\documentclass[11pt]{article}
\usepackage{graphicx} 
\usepackage{amssymb}
\usepackage{authblk} 
\usepackage{hyperref} 

\usepackage{amssymb} 
\usepackage{amsmath} 

\usepackage{geometry}

\usepackage{xcolor}

\title{Difficult control is related to instability in biologically inspired Boolean networks}

\author[1*]{Bryan C. Daniels}
\author[1**]{Enrico Borriello} 
\usepackage{url}

\affil[1]{School of Complex Adaptive Systems, Arizona State University, Tempe, AZ, USA} 
\affil[*]{\small\url{bryan.daniels.1@asu.edu}}
\affil[**]{\small\url{enrico.borriello@asu.edu}}

\date{\today}

\begin{document}

\maketitle

{\small {\bf Abstract} 
Previous work in Boolean dynamical networks has suggested that the number of components that must be controlled to select an existing attractor is typically set by the number of attractors admitted by the dynamics, with no dependence on the size of the network.  Here we study  the rare cases of networks that defy this expectation, with attractors that require controlling most nodes.  We find empirically that unstable fixed points are the primary recurring characteristic of networks that prove more difficult to control.  We describe an efficient way to identify unstable fixed points and show that, in both existing biological models and ensembles of random dynamics, we can better explain the variance of control kernel sizes by incorporating the prevalence of unstable fixed points. In the end, the association of these outliers with dynamics that are unstable to small perturbations reveals them as artifacts of deterministic models, making them less biologically relevant and reinforcing the generality of easy controllability in biological networks.} 

\section{Introduction}


The formulation of an exhaustive theory for controlling complex biological systems, and gene regulatory networks (GRNs) more specifically, remains a key objective in systems biology \cite{kitano2002computational,richardson2015postgenomics}. And while optimal control theory offers a broad set of tools for various applications in mathematics and engineering, its traditional assumptions often require modification for biological contexts. Biological processes are typically inherently nonlinear \cite{isidori2013nonlinear}, and the focus is frequently restricted to coarse-grained output states \cite{borriello2020cell}, rather than driving the system to an arbitrary state \cite{ZanYanAlb17}.
There are a number of approaches to quantifying control in biology,
which vary in their basic assumptions.  States can be modeled as continuous or discrete, dynamics can be assumed to be stochastic or deterministic, control inputs can be static or dynamic, and the goal of control can be to reach arbitrary states or to reach particular phenotypes (attractors). Much work in biological control has focused on phenotypic control toward specific coarse-grained output states \cite{borriello2020cell},  as it is arguably more relevant to biological function than driving a system into any arbitrary state \cite{ZanYanAlb17}.  Some work inspired by traditional control theory has focused on control toward arbitrary states, either in continuous \cite{liu2011controllability} or discrete state space \cite{CANA,ElSKha10}, though this typically leads to control strategies that are significantly more difficult to implement than what is necessary to select a single desired output.
The simplicity of discrete state spaces (and the inability to constrain parameters in continuous formulations) has guided many to take a Boolean approach. The typical interventions available in experiments has led to a focus on control inputs that are largely static, either applied permanently as in a  gene knockout or temporarily over some timescale \cite{ActoNet,AEON, CABEAN,Caspo,pystablemotifs}, though some approaches allow for arbitrary dynamics of inputs \cite{CANA,CABEAN}. Finally, for simplicity and mathematical tractability, the updating functions of Boolean network models are often 
defined deterministically.   A mild form of stochasticity is sometimes incorporated in defining the order in which nodes are updated across time (e.g., \cite{ActoNet,AEON,CABEAN,pystablemotifs}), or  noise can be modeled in the form of bit-flips or probabilistic transitions \cite{MurVelAgu12}.

We focus here on discrete, Boolean, deterministic, static, phenotypic control.
In this context, an important concept for regulating network dynamics is the {\it control kernel} (CK). First proposed in \cite{kim2013discovery}, the CK refers to a set of genes of minimal size whose external manipulation is enough to guide the network towards a target steady-state gene activation pattern, i.e. an attractor of the dynamical model of the GRN.  The CK is defined such that it can force a particular attractor state regardless of the initial condition. This matches with the assumption that biological function is mostly captured by attractors, in that they represent phenotypically and functionally distinct cell types \cite{Kauffman1969}.

Several related approaches have been proposed for the study of the optimal control of deterministic Boolean networks. Stable motif analysis defines control sets by detecting positive circuits within the network that can sustain trap spaces in the dynamics \cite{ZanAlb15}. Feedback vertex sets offer an efficient upper bound on the size of minimal control sets in a way that does not depend on the specific dynamics governing each node \cite{FieMocKur13,MocFieKur13,ZanYanAlb17}.




A simple heuristic argument suggests that the size of CKs may scale approximately with the logarithm of the number of original attractors. Under the simplifying assumption that state transitions are all equally likely, forcing a single node into a specific state would generally halve the number of attainable states. Therefore, by controlling $c$ nodes, we would expect to reduce $2^c$ possible attractors down to a single attractor, suggesting the logarithmic scaling of CK size.
However, the task of identifying the minimal set of controlling nodes is nontrivial and has been proven NP-hard by Akutsu {\it et al.} \cite{akutsu2007control}, and counterexamples can be constructed that require much larger CKs. 

In Ref.~\cite{borriello2021basis}, we demonstrated that dynamic Boolean networks do often have CK sizes that scale logarithmically with the number of attractors. We illustrated this on a large database of experimentally derived biological networks, as well as ensembles of random networks.  These previous findings revealed that control toward existing attractors does not inherently become more difficult as the size of the network increases.  Instead, the number of attractor states generated by the dynamics is a much better predictor of the difficulty of control than the network size.  Our result could then be summarized in the statement that the average size of the control kernel of a dynamical Boolean network, $ \langle |\textrm{CK}| \rangle$, typically scales logarithmically with the number of its attractors, $r$:
\begin{equation}
    \langle |\textrm{CK}| \rangle \simeq   \log_2{r}.
    \label{eq:scaling}
\end{equation}
In Ref.~\cite{borriello2021basis}, we reinterpreted this result in terms of the conjectured expectation value of the witness set in computational learning theory \cite{kushilevitz1996witness}. More importantly, we view it as one of the most significant theoretical and empirical justifications to date for the feasibility of genetic reprogramming \cite{MulSch11}.

Unbeknownst to us, research conducted by Akutsu's group and published earlier than our work had reached a very similar conclusion and independently identified the logarithmic scaling described, even if under somewhat different assumptions and on more abstract models \cite{hou2019number}.
A number of clever simplifications make the mathematics more manageable and allow for the direct derivation of logarithmic scaling as in Eq.~\ref{eq:scaling}.
The primary difference between the control defined by Hou et al.\ and our study is the duration of the pinning. Hou et al.\ assume that the pinning, which constitutes the control signal, lasts for a single timestep and is then removed, whereas we assume that the pinning is permanent. 
Our choice of permanent pinning is influenced by our focus on biological networks, where long-timescale perturbations, such as those observed in gene overexpression, knockdown, and knockout experiments, have practical significance. 
Another important difference is that Hou et al.\ assume that the basins of all attractors are of equal size. Due to the large variability in the basin sizes of the experimentally motivated models we examine, our current work aims to shed more light on the still hazy relationship between basin size and controllability.

In our previous study, some networks were much more difficult to control than predicted by a logarithmic scaling law (roughly 2\% of tested networks).
A few networks had significant numbers of attractors whose control required pinning nearly all nodes in the network.  The presence of such networks, along with the unaccounted reasons for their difference from more easily controllable networks, could undermine one's confidence in predicting the required level of control based solely on knowledge of a network's attractors, as well as the hypothesis that a network's size is typically irrelevant to the difficulty of control.

Our focus in this work is to investigate the factors that contribute to the increased difficulty of controlling such networks. Our goal is to find approximations to the mean control kernel size that are simple to compute and do not rely on solving the full dynamics, which becomes intractable in moderately sized networks.

As a first step, we expand our database of biological networks by incorporating additional networks collected in Ref.~\cite{kadelka2020meta}, representing the most extensive collection of Boolean models for biological networks at the time of writing this manuscript. None of the new networks exhibit a significant deviation from our original predictions, leaving us with the task of explaining only the outliers previously identified.

A direct examination of these outlier networks highlights the presence of isolated fixed points, i.e., fixed points of the dynamics that do not attract any other state. We establish a strong correlation between this characteristic and the difficulty of control, and we use this as the rationale for a correction to our original scaling law (Equation \ref{eq:scaling}). The revised scaling law relies not only on the knowledge of the number of attractors but also on the number of isolated fixed points among them. Interestingly, the latter can be easily evaluated even for networks where not all basin sizes are known, particularly in cases where the identity of all attractors can be obtained by exploiting  modularity in the network topology. We test whether our new empirical formula outperforms our previous scaling law, and demonstrate that it accounts for the majority of the remaining variance not captured by the original scaling.

In the next section ({\it Results}), we describe our extended analysis and provide the correction to our empirical formula.

In the {\it Methods} section, 
we show how we leverage the modular structure of biological networks to identify the complete spectrum of their attractor states, and we demonstrate how to exploit the sparse connectivity of these networks to identify all isolated fixed-points by explicitly determining their backward reachable sets, i.e., the sets of their pre-images.

Finally, in the {\it Discussion} section, we comment on the biological relevance of these isolated fixed-points.

\section{Results}

To explore the drivers of unusually large control kernels, we first ask: What is the maximum possible control kernel size for a given network?  
Given our definition of control, fixed-point attractors always have a control kernel, as control can be attained by pinning all nodes in the network.
Note that cyclic attractors are sometimes not able to be forced with static control because we are limited to pinning non-cycling nodes --- in these cases, we say that a control kernel does not exist.
This sets the simplest upper bound on the size of control kernels for controllable attractors of $|\mathrm{CK}| \leq n$.  A more economical approach avoids pinning nodes in any peripheral trees that are wholly dependent on nodes within closed feedback loops in the regulatory network topology. We will refer to the network's core as the set of nodes that exclude any peripheral trees, with a number of core nodes $n_\mathrm{c}$. As control kernels are defined as sets of controlling nodes of minimal size, we can ensure a tighter bound of $|\mathrm{CK}| \leq n_\mathrm{c}$. 

We might naively expect that it would be more difficult to steer the system toward attractors with a smaller than average basin size when initializing the system in a random state.
Therefore, one might expect extreme control kernel sizes for attractors with very small basins of attraction (defined as the states that lead to each attractor under the dynamics). While basin sizes are not generally easy to compute, it is relatively simple to highlight the most extreme scenarios when an attractor is an ``isolated'' fixed point, i.e., with a basin size of 1 (see Methods). 

\begin{figure}[tbp]
    \centering
    \includegraphics[width=0.95\linewidth]{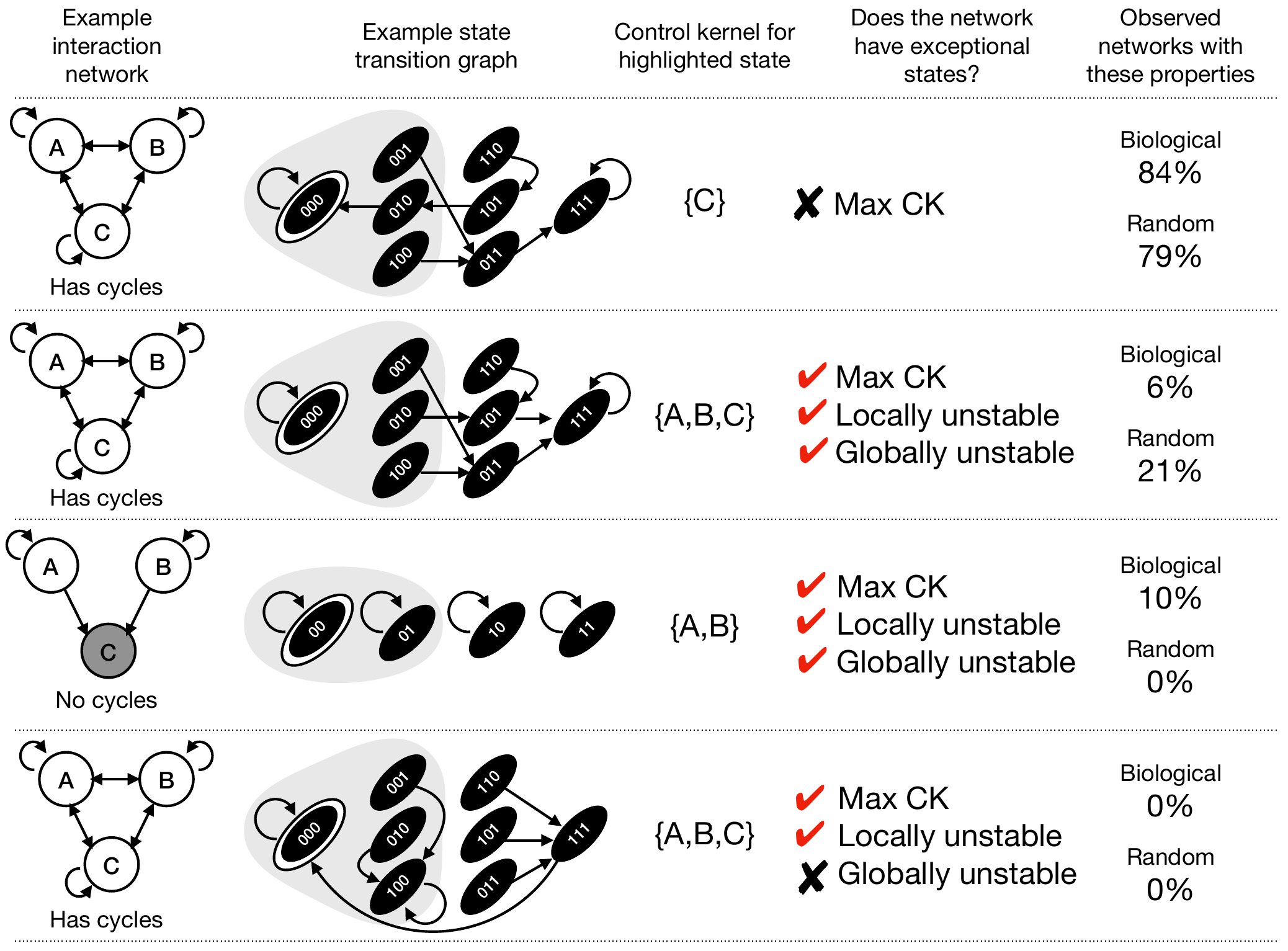}
    \caption{\textbf{Examples of networks with and without exceptional states.}
    Networks that have control kernels of maximal size $n_c$ (``Max CK'') 
    are related to those that have locally and globally unstable states.  
    Empirically, most networks do not have control kernels of maximal size
    (first row).
    All observed networks that we identify as being unusually 
    difficult to control contain both fixed points 
    with maximal CK size and globally unstable fixed points (second row).  
    Some of the biological networks have no cycles in their interaction networks,
    which trivially creates $2^m$ fixed points 
    with $|\mathrm{CK}| = m = n_c$; since they follow Eq.~\ref{eq:scaling},
    we do not consider these networks to be 
    unusually difficult to control (third row).
    Finally, it is possible to construct networks that contain fixed points with
    maximal CK size but no globally unstable fixed points, but we find no
    examples of this case in the sampled networks (fourth row).
    Globally unstable fixed points have an incoming edge only from themselves
    in the state transition graph.  Locally unstable fixed points can have
    other incoming edges, but none from nearest Hamming neighbors (gray region).
    Fixed points with maximal CK size must be locally unstable.
    }
    \label{fig:example-networks-table}
\end{figure}

\begin{figure}[tbp]
    \centering
    \includegraphics[width=0.5\linewidth]{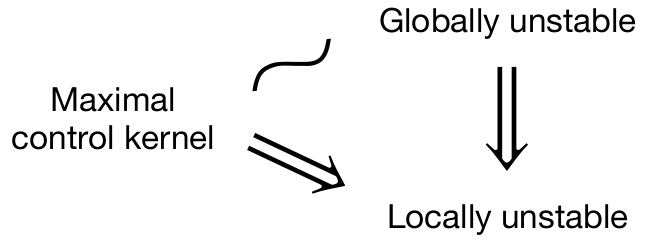}
    \caption{\textbf{Summary of the logical relationships between the stability of a fixed point and whether it has a control kernel of maximal size.}
    A fixed point with maximal control kernel size ($|\mathrm{CK}| = n_\mathrm{c}$) is always locally unstable, as is one that is globally unstable.  Global instability is correlated with maximal control kernel size in the ensembles we study, but neither implies the other in general.  Here local instability means that all states at Hamming distance 1 are in the basins of other attractors, and global instability means that no other states lead to the fixed point (basin size $= 1$).
    }
    \label{fig:stability-schematic}
\end{figure}

Both isolated fixed points and maximal control kernel sizes ($|\mathrm{CK}| = n_\mathrm{c}$) are connected to concepts of instability (examples shown in Fig.~\ref{fig:example-networks-table} and summarized in Fig.~\ref{fig:stability-schematic}).  First, fixed points with maximal control kernels are locally unstable, in the sense that such a fixed point must be unstable to individual bit flips.  That is, each state at Hamming distance 1 from the fixed point must be in the basin of a different attractor (otherwise one of the nodes could be left unpinned and the system would still return to the fixed point).  Note that local instability does not necessarily imply that no other states transition into the fixed point under the dynamics (a counterexample is shown in the last row of Fig.~\ref{fig:example-networks-table}).  Also, while having a maximal control kernel implies local instability, local instability does not in itself guarantee a maximal control kernel.  Second, isolated fixed points are globally unstable, in the sense that no other states transition into them.\footnote{This is distinct from other notions of instability in the literature on Boolean networks, such as instability of cycles to delays in updating \cite{KleBor05}.}
Though a fixed point being globally unstable does not directly imply that it has a maximal control kernel, we might expect some correlation between instability and difficulty of control.

\begin{figure}[tbp]
    \centering
    \includegraphics[width=\linewidth]{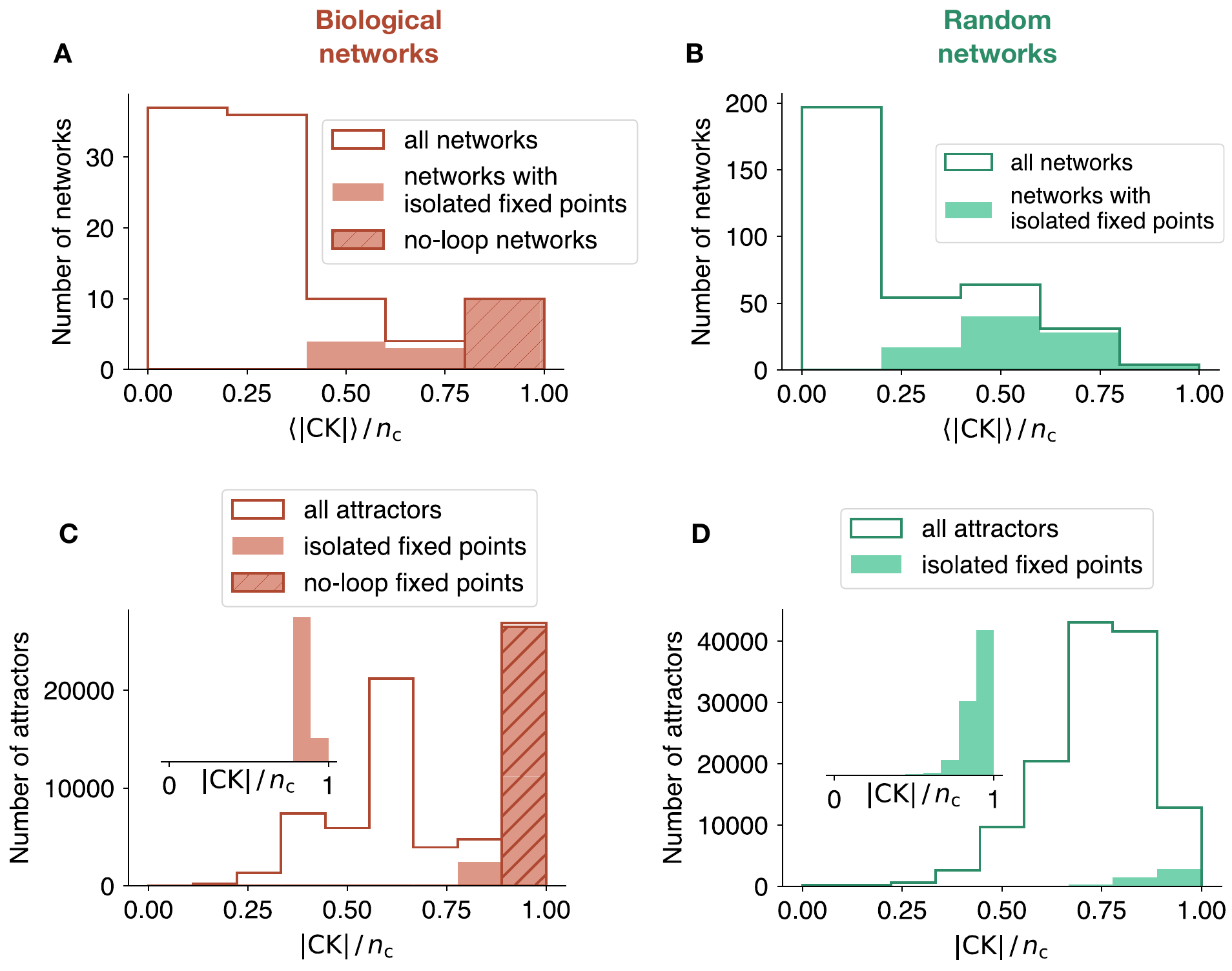}
    \caption{\textbf{Isolated fixed points are related to large control kernel sizes.} Both when averaged across attractors within networks (A and B) and at the level of individual attractors (C and D), isolated fixed points (those with basin size = 1; shaded bars) produce control kernel sizes $|\mathrm{CK}|$ that are large relative to the total number of core nodes $n_\mathrm{c}$.  This pattern holds in both the biological (A and C) and random (B and D) network ensembles that we study.  Some of the biological networks have no feedback loops, which trivially creates isolated fixed points with $|\mathrm{CK}| = n_\mathrm{c}$ (hatched bars in A and C).  Insets in C and D highlight the distributions of control kernel sizes for only the isolated fixed points in networks that do contain feedback loops.}
    \label{fig:isolated-and-nonisolated-ck-size-histograms}
\end{figure}

In Fig.~\ref{fig:isolated-and-nonisolated-ck-size-histograms}, we compare average control kernel sizes in networks that have isolated (globally unstable) fixed points to the distribution over all networks (A and B) and control kernel sizes for individual isolated fixed points compared to the distribution over all attractors (C and D).  The control kernel sizes of isolated fixed points are often close to $n_\mathrm{c}$, matching with our intuition.  Some of the biological networks have no feedback loops and are therefore controlled entirely by their input nodes, which trivially creates isolated fixed points with $|\mathrm{CK}| = n_\mathrm{c}$ (hatched areas in Fig.~\ref{fig:isolated-and-nonisolated-ck-size-histograms} and third row in Fig.~\ref{fig:example-networks-table}). 
Even aside from this effect, isolated fixed points tend to have very large control kernels.  

\begin{table}
    \centering
    \begin{tabular}{cccccc}
         & Total & Controllable & With Isolated & With Max CK & With Both \\
        Biological networks & 104 & 97 & 17 & 16 & 16 \\
        Random networks & 371 & 346 & 87 & 73 & 73 \\
    \end{tabular}
    \caption{\textbf{Summary of the analyzed networks.} ``Total'' includes all networks for which we can compute attractors and control kernels exactly. ``Controllable'' includes all networks for which static control kernels exist for a non-negligible set of attractors (here set as a threshold that the basins of uncontrollable cycles must make up less than 99\% of the state space). ``With Isolated'' includes networks that have at least one isolated fixed point.  ``With Max CK'' includes networks that have at least one fixed point with $|\mathrm{CK}| = n_\mathrm{c}$.  ``With Both'' includes networks that have nonzero numbers of both isolated fixed points and maximal control kernels.  (Note: 10 of the biological networks have no loops, which implies that all of their attractors are isolated and control kernels are of maximal size.)}
    \label{table:network-counts}
\end{table}

Thus, though global instability does not always imply difficult control, the correlation is nonetheless striking in the ensembles that we study.  In fact, we find in our network ensembles that fixed points with maximal 
control kernels occur only in networks that have at least one globally unstable fixed point (Table~\ref{table:network-counts}).

At the level of individual networks and attractors there are exceptions to the rule that globally unstable fixed points tend to be difficult to control. See SI Fig.~\ref{fig:individual-network-histograms-random} and Fig.~\ref{fig:individual-network-histograms-biological} for histograms of control kernel sizes within each network that has at least one globally unstable fixed point.  (In one network from the random ensemble, a globally unstable fixed point in fact has the smallest control kernel of all the network's attractors.)  Yet these exceptions are rare.

\begin{figure}[tbp]
    \centering
    \includegraphics[width=\linewidth]{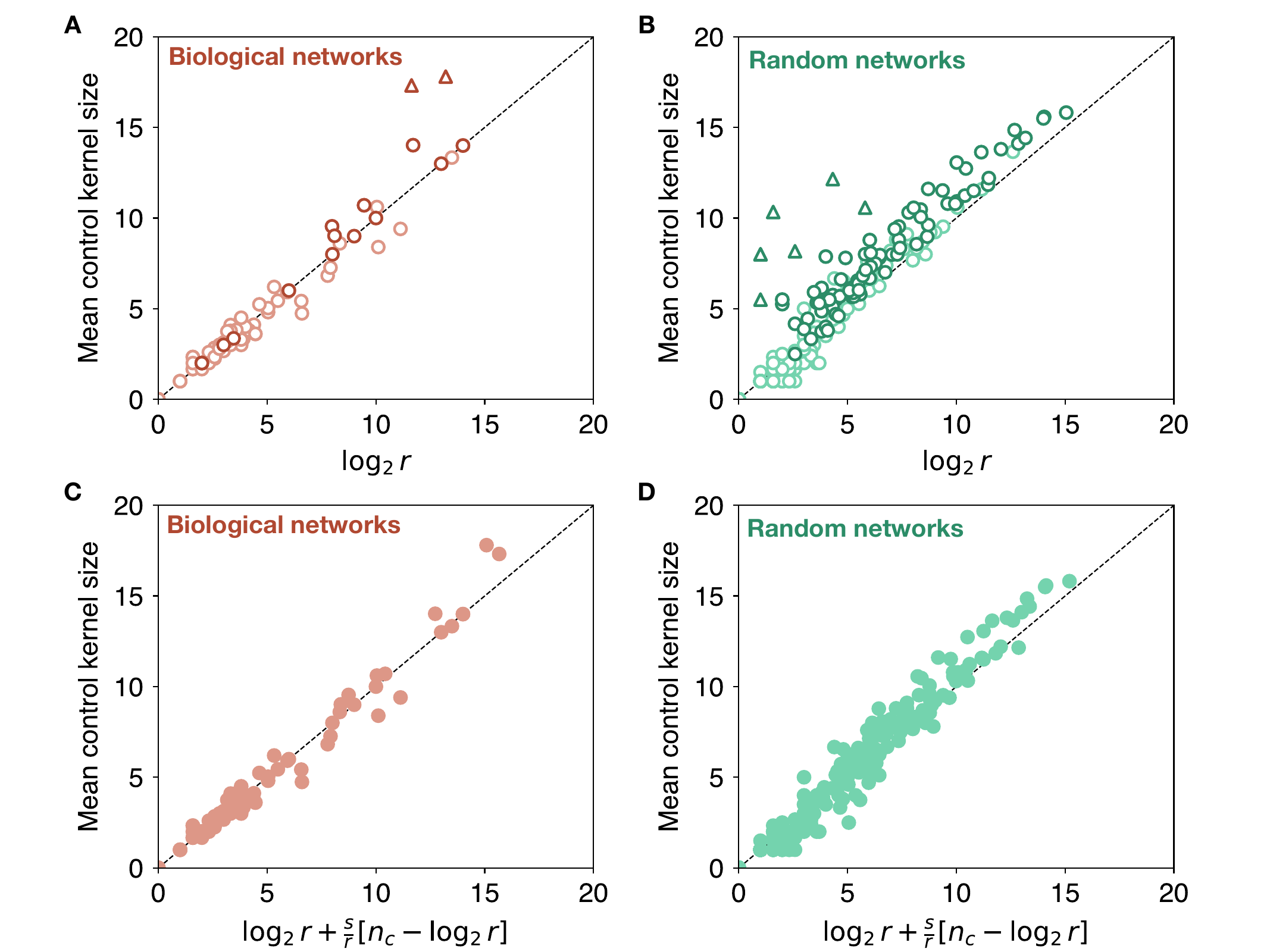}
    \caption{\textbf{A correction that incorporates the number of isolated fixed points leads to better predictions of control kernel sizes.} (A and B) The simple prediction that the mean control kernel size within each network is equal to the logarithm of the number of attractors $r$ (Eq.~\ref{eq:scaling}) works well for most networks, but there are a number of outlier networks that are much harder to control than expected.
    We mark networks as outliers (triangles) when their mean control kernel size
    is more than $3\sigma$ away from the expected value of $\log_2{r}$, where
    $\sigma$ is the standard deviation of the residuals $\langle |\textrm{CK}| \rangle - \log_2{r}$.
    These outlier networks all have isolated fixed points (highlighted with a darker color).
    (C and D) A correction that assumes isolated fixed points have control kernel of size $n_\mathrm{c}$ (Eq.~\ref{eq:scaling-corrected}) leads to better predictions.  For the biological networks, the RMSE is 0.93 when using
    Eq.~\ref{eq:scaling} and reduces to 0.60 when using Eq.~\ref{eq:scaling-corrected}.  For the random networks, the RMSE reduces from 1.3 to 0.68. }
    \label{fig:ck-size-predictions}
\end{figure} 

Given that it is relatively fast to determine whether a given fixed point is globally unstable (see Methods), and these are also  the ones that tend to contribute most to changes in $\langle |\mathrm{CK}| \rangle$, we propose using their number to get a computationally simpler estimate of control kernel sizes.  
If we assume that isolated fixed points are the only ones contributing to the difference from Eq.~\ref{eq:scaling} and that they have the maximum possible size $n_\mathrm{c}$, we get the approximation
\begin{equation}
\langle |\textrm{CK}| \rangle \simeq \frac{r-s}{r} \log_2{r} + \frac{s}{r} n_\mathrm{c} \ ,
\label{eq:scaling-corrected}
\end{equation}
where $s$ is the number of isolated fixed-points.  As shown in Fig.~\ref{fig:ck-size-predictions}, this simple estimate does quite well in most cases, outperforming Eq.~\ref{eq:scaling} for almost all networks we tested.
In particular, if we define networks that are unusually difficult to control
as those that are outliers in Fig.~\ref{fig:ck-size-predictions}A and B
(shown as triangles, and defined as those with residuals $|\langle |\textrm{CK}| \rangle - \log_2{r}| > 3\sigma$, with 
$\sigma$ the standard deviation of those residuals across networks in each
ensemble),
Eq.~\ref{eq:scaling-corrected} successfully corrects the predictions 
of mean control kernel sizes for those difficult cases.

The significance of Eq.~\ref{eq:scaling-corrected} lies in the fact that, similar to Eq.~\ref{eq:scaling}, it enables an estimation of the average control kernel size solely based on the characteristics of the attractor landscape of the dynamics. This is achieved without the necessity of determining control kernels for all $r$ attractors. 

\begin{figure}
    \centering
    \includegraphics[width=0.55\linewidth]{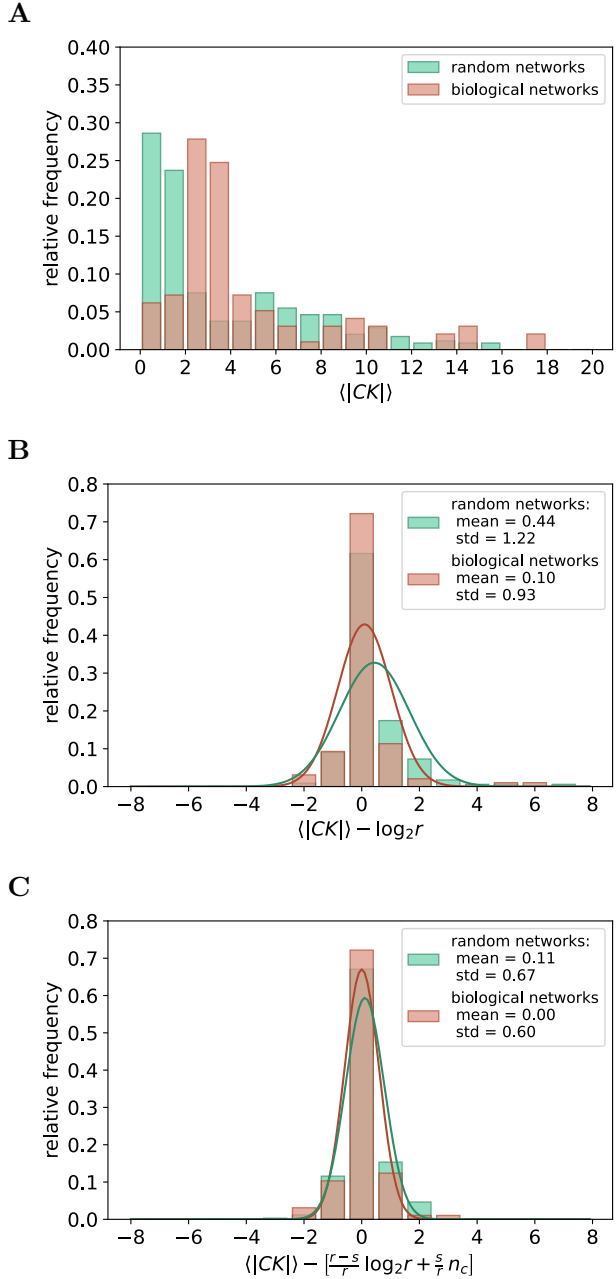}
    \caption{\textbf{The isolated fixed point correction reduces the variance of mean control kernel size predictions.} 
    (A) The distributions of mean control kernel sizes across the ensembles of biological and random networks have large variances (listed in the legend).
    (B) Most of this variance can be explained in terms of the number of attractors $r$ as in Eq.~\ref{eq:scaling}.
    (C) A large fraction of the remaining variance is explained by using the corrected Eq.~\ref{eq:scaling-corrected} that incorporates the number of isolated fixed points $s$.}
    \label{fig:variance-histograms}
\end{figure}

Fig.~\ref{fig:variance-histograms} illustrates the initial variance in the distribution of $\langle|\mathrm{CK}|\rangle$ across the analyzed biological and random networks. This variance can be better explained by our new predictor in Eq.~\ref{eq:scaling-corrected} as compared to the previous predictor in Eq.~\ref{eq:scaling}.
The improved precision in the new estimation is attained by integrating additional information, specifically the number $s$ of isolated fixed points within the landscape, which is the primary cause for the deviations from our simple estimate. 

Significantly, Eq.~\ref{eq:scaling-corrected} introduces a  dependence of the control kernel on the network size (specifically on $n_\mathrm{c}$). This seems to be at odds with our findings in Ref.~\cite{borriello2021basis}, where we concluded that the average control kernel size remains independent of the network size for a given $r$. It is crucial to note that the parameter $s$ is typically small in comparison to $r$, so that the influence of $n_\mathrm{c}$ is often hidden upon considering an average across all attractors in a network. This becomes especially apparent when Eq.~\ref{eq:scaling-corrected} is recast in the form
\begin{equation}
\langle |\textrm{CK}| \rangle \simeq \log_2{r} + \frac{s}{r} \Big[ n_\mathrm{c}  - \log_2{r} \Big] \ ,
\label{eq:scaling-corrected-rearranged}
\end{equation}
which shows how the size-dependent contribution is a correction to Eq. \ref{eq:scaling} whose relevance scales with the fraction $s/r$ of isolated fixed-points.

Once analyzed in the context of their biological relevance, we may expect the dependence on $n_\mathrm{c}$ to be even less impactful in real-world dynamics. The presence of isolated fixed-point attractors, which arise in models with deterministic updating rules, is undermined by the incorporation of stochasticity in more realistic models of gene regulation (see {\it Discussion}). 

Finally, treating Eqs.~\ref{eq:scaling} and \ref{eq:scaling-corrected} as
statistical predictors of average control kernel size, 
we examine the distributions of residuals away from these predictions.
The non-normality of the residual distribution for the uncorrected 
Eq.~\ref{eq:scaling} is visible in Fig.~\ref{fig:variance-histograms}B,
with the outlier networks contributing to a sizable skew.
This gives a statistical motivation to look for a better predictor.
Our corrected Eq.~\ref{eq:scaling-corrected} then produces a distribution
of residuals that is closer to normal, with the outliers largely 
gone (Fig.~\ref{fig:variance-histograms}C).  
We quantify this approach to
normality using the Shapiro-Wilk statistic: for the biological networks, 
$W$ increases from $0.64$ using Eq.~\ref{eq:scaling} to $0.85$ using 
Eq.~\ref{eq:scaling-corrected}, and for the random networks, 
$W$ increases from $0.68$ to $0.87$.  See SI section \ref{section:residuals} and 
Fig.~\ref{fig:q-q-plots} for a more thorough discussion and visualization 
of the residual distributions.

\section{Methods}

\noindent
{\bf Boolean Networks and attractor dynamics.}
In this section, we define key terms and concepts related to Boolean dynamical systems. Readers already familiar with Boolean network dynamics may skip to the next section.


The state of a Boolean network at time $t$ is characterized by the states of its $n$ nodes, $x_i(t)$, where $i=1,\dots,n$. Due to the Boolean nature of the model, each node $i$ can be in one of two possible states: ON, represented by $x_i(t)=1$, or OFF, represented by $x_i(t)=0$. In the context of cell regulation, the nodes typically correspond to interacting genes, with $x_i(t)$ serving as the Boolean approximation of the expression level of gene $i$. Here, the gene is considered expressed when $x_i(t)=1$, and inactive otherwise.

The dynamics of a deterministic Boolean network are characterized by $n$ Boolean functions $f_1,\dots,f_n$, which take as input the array $x_1(t),\dots,x_n(t)$ and produce the network's configuration at time $t+1$:
\begin{eqnarray} 
x_1 (t+1)  & = & f_1 (x_1(t),\dots,x_n(t)) \nonumber \\
\cdots \ \ \qquad &   &                  \nonumber \\
x_n  (t+1) & = & f_n (x_1(t),\dots,x_n(t)) \ . \label{system1_alternate} 
\end{eqnarray}
The specific form of these Boolean functions can be identified through experiments \cite{Henry2013,Zhou2016}. This setup represents a {\it synchronous} update of all nodes in the network. 

With $n$ nodes, there are $2^n$ potential network configurations (i.e., gene expression patterns in the context of a genetic network). We will refer to these configurations as {\it states} of the network and the collection of all possible network states as the {\it configuration space}. The dynamics of the network are depicted by a time series of states. Although large, the configuration space is finite. Hence, when the functions $f_i$ are deterministic, these dynamical trajectories will eventually settle into either a fixed state or a cycle of states. These specific sets of states are termed the {\it attractors} of the Boolean network. The collection of initial states that converge to a particular attractor is called its {\it basin of attraction}. 


\vspace{4mm}\noindent{\bf Control kernels.}
We characterize the difficulty of control using the notion of a control kernel \cite{kim2013discovery}.  Specifically, we define a control kernel for a given attractor as a set of nodes of minimal size that can be statically pinned to fixed values such that, in the pinned dynamics, all initial conditions converge to the desired attractor.  Using this static definition of control, all fixed point attractors have a control kernel of some size, but not all cycles are controllable.

Many biological network models include nodes that represent external or environmental conditions.  We call such nodes ``input nodes,'' and we implement their dynamics using the updating rule $x(t+1) = x(t)$. Input nodes are included in the network state, and we incorporate all possible values of these nodes when enumerating a network's attractors. Based on our definition of control, all input nodes are included in every control kernel, as no other nodes can influence their state \cite{borriello2021basis}. In rare cases, a network model will define a node's dynamics such that it always results in a specific constant value $C$, that is, $x(t+1) = C$, with $C = 0$ or $1$.  Such nodes have a fixed value across all possible attractors. They therefore do not distinguish different attractors and are thus never included as part of a control kernel.

For set environmental conditions (fixed states of input nodes) 
multiple attractors can only exist if the network structure presents feedback loops \cite{zanudo2017structure}. Any set of nodes connected in a tree structure without closed loops will depend deterministically on the states of input nodes in each attractor state.  For this reason, nodes outside of a feedback loop will never be part of a control kernel (as all behavior of trees will be fully controlled once the attractor is selected within the core).  For this reason, we remove from our analysis all dependent trees of nodes from each network, leaving only a number $n_\mathrm{c} \leq n$ of ``core'' nodes that participate in feedback loops and have the potential to be part of control kernels.\footnote{
    A subtlety arises in computing $n_c$ in that redundant expressions in the updating rules can create inoperative edges \cite{GatCorWan21}.  We detail in supplementary section \ref{section:redundant} why we do not expect this to affect our results.
}  Alternatively, we could include all $n$ nodes in our analysis, but this would preclude the possibility of having any isolated fixed points in networks that have any nodes on which no other nodes depend (as changing the states of such nodes does not affect the dynamics).  

Finally, we note that the core nodes of a network with no cycles consist only of the network's input nodes. This corresponds to the fact that every attractor is fully specified by the states of the inputs, and results in the control kernel being equal to the set of input nodes.  It is also possible for inputs to fully control networks that include cycles.  In all of these cases, the number of attractors is $2^m$, control kernels are of size $m$, and the logarithmic scaling of Eq.~(\ref{eq:scaling}) is satisfied exactly.  Therefore, these cases do not show up among those networks that we consider unusually difficult to control.

\vspace{4mm}\noindent{\bf Network ensembles.}
For this study, we complement the set of network models analyzed in Ref.~\cite{borriello2021basis} with the more recent database compiled by Kadelka et al. \cite{kadelka2020meta} consisting of 122 Boolean models of biological regulation. The models in this database were chosen from a pool of 163 models identified using the Pubmed biomedical literature search engine, further restricted to include only expert-curated models (where both nodes and update rules were manually selected to prevent artifacts induced by prediction algorithms) and only one version of closely related models. Out of the 122 models, 61 were present in the Cell Collective database \cite{Helikar2012} that we analyzed in Ref.~\cite{borriello2021basis}. For consistency with our previous work, we retain a small number of Cell Collective networks that were not included in the Kadelka et al.\ database.  
The resulting set of networks encompasses models describing the regulatory logic governing various processes across a diverse array of species spanning different kingdoms of life, including animals, plants, fungi, and bacteria. For additional details, see Ref.~\cite{kadelka2020meta} and its supplementary information.

A key limitation of Boolean dynamical models is the need to limit their size, as the number of network states grows exponentially with the number of nodes. To manage this, peripheral regions of the interactome are often simplified into input and output nodes, based on the assumption that the highlighted subsystems function relatively independently of the complete cellular interactome. While studies have shown that the network topology of these reduced models partially preserves the motif distribution of the full networks \cite{borriello2024local}, unsurprisingly this preservation diminishes as the size of the extracted subgraphs decreases.

To increase the variance in our tested models, we also include three ensembles of random Boolean networks, as presented in Ref.~\cite{borriello2021basis}. These include networks with random truth tables, as well as Erd{\"o}s-R{\'e}nyi networks governed by threshold rules, with thresholds chosen either at zero total input or balanced to account for the number of inputs to each node. 
First, we use the well-studied ensemble of p-K random
networks \cite{ShmKau04} to generate dynamics with random truth tables.  
Each node receives input from $K=2$ nodes, with Boolean updating rules 
chosen such that the probability of the ON state is $p$.  
We sampled 225 networks from this ensemble, 
with 75 each having $p = 0.25$, $0.5$, and $0.75$, 
and 25 each within these sets having number of nodes $n=10$, $15$, and $20$.  
We found control kernels for all controllable attractors 
in each of these sampled networks.
Second, we define an ensemble that 
assigns each node a set of input nodes whose states are summed and 
compared to a threshold \cite{LiLonLu04,kim2013discovery}.
The network's dependency structure $A$ is chosen as an Erd{\"o}s-R{\'e}nyi graph with average degree $d$, and each edge in this graph is assigned with probability $p_I$ a value of $-1$ (representing an inhibitory interaction) or $+1$ (representing an excitatory interaction). Each node's state is determined by comparing the sum of the incoming signed inputs $s_i = \sum_j A_{ij} x_j(t)$ to its threshold $\tau_i$:
\begin{equation}
x_i(t+1) = 
\begin{cases}
0, \textrm{if}~s_i < \tau_i \\
x_i(t), \textrm{if}~s_i = \tau_i \\
1, \textrm{if}~s_i > \tau_i.
\end{cases}
\end{equation}
For one ensemble, we set thresholds to zero ($\tau_i = 0~\forall~i$).   
In cases with little inhibition,
these networks are biased toward excitation.
A second ensemble consists of ``balanced'' threshold networks, 
with thresholds of individual nodes set at the center of the 
distribution of possible summed inputs ($\tau_i = \frac{1}{2} \sum_j A_{ij}$).
We sampled 75 networks using each type of threshold, 
with each combination of $d = \{1,2,3,4,5\}$, $p_I = \{0.1,0.3,0.5,0.7,0.9\}$, 
and $n = \{10, 15, 20\}$, for a total of 150 networks.  
Of these, for 146 we were successful in finding control kernels for all controllable attractors.

The control results for all random networks included in this study are shown in Fig.~\ref{fig:ck-size-random-separate}, where we replot Fig.~\ref{fig:ck-size-predictions}B to highlight the differences arising from the various methods used to construct the random networks. Notably, all outliers, marked as triangles in Fig.~\ref{fig:ck-size-predictions}B, originate from the zero-threshold ensemble.

We use a slightly stricter criterion for choosing networks to include here than  in previous work \cite{borriello2021basis}. As before, we only include networks for which attractors and control kernels can be computed with a reasonable amount of computer time (about 1 day of compute time on a single CPU), and we further restrict ourselves to include only networks for which we are able to compute attractors exactly, without relying on sampling. We are able to compute exact attractors and control kernels for 104 biological networks (out of 122) and 371 random networks (out of 375).
Furthermore, a limited number of networks (7 out of 104 biological networks and 25 out of 371 random networks) are characterized by uncontrollable cycles with basins that cumulatively comprise more than 99\% of the state space. Due to the static nature of the control kernel examined in this study, we filter out these networks from further analysis. 

Complete lists of networks analyzed in this work are included as comma separated files in the supplementary material, and the biological networks are listed in Tables \ref{tab:cc_nets} and \ref{tab:iowa_nets}.

\vspace{4mm}\noindent{\bf Modularity.} 
Biological regulatory network models often exhibit high modularity in their topology, where subsets of nodes interact more intensely within their own subsets than with other subsets. Additionally, these models tend to be hierarchical, featuring upstream modules that operate independently of downstream module behavior \cite{paul2018decomposition}.

Efficient computation of attractors in modular and hierarchical Boolean networks can be achieved by analyzing modules separately. The approach involves identifying attractors for upstream modules first and subsequently leveraging each upstream attractor to determine corresponding attractors for downstream modules. Despite some intricacies discussed below, the computational execution of this process remains straightforward.

We first decompose a given Boolean dynamical system into hierarchical modules, defined as strongly connected components within the system's causal network. This representation is in the form of a directed acyclic graph, capturing dependencies among modules. We then identify attractors for upstream modules that are independent of others, utilizing a brute-force approach iterating over all possible states within the module.

For each downstream module $\mathcal{D}$, attractors $\textbf{U}$ of upstream modules, on which $\mathcal{D}$ relies, are explored through all possible combinations. Given an attractor $U \in \textbf{U}$, the states of nodes in upstream modules are fixed to the values present in $U$, and corresponding attractors for $\mathcal{D}$ are determined using a brute-force approach iterating over all possible dynamics of $\mathcal{D}$ based on the input from $U$.

Two subtleties arise when computing the set of all possible attractors $\textbf{U}$ for nodes upstream of $\mathcal{D}$. In the simplest case, where $\mathcal{D}$ depends on separate modules $\mathcal{U}_1, \mathcal{U}_2, ..., \mathcal{U}_n$ without shared ancestor modules, and all corresponding sets of upstream attractors $\textbf{U}_1, \textbf{U}_2, ..., \textbf{U}_n$ consist only of fixed points, $\textbf{U}$ comprises $\prod_i |\textbf{U}_i|$ attractors. However, when upstream modules share ancestor modules further upstream, or when dealing with cyclic attractors of length $\ell > 1$, additional considerations are necessary to account for inconsistent combinations and phase shifts between attractors, respectively.

We use this modular approach to efficiently identify attractors and compute control kernels.  For more details, see Ref.~\cite{borriello2021basis}.

\vspace{4mm}\noindent{\bf Backward Reachable Sets.} 
Equation \ref{eq:scaling-corrected} relies on evaluating the basin sizes, a non-trivial task. To find basin sizes, one must solve for the equilibrium dynamics starting from every state of the system. The size of the network becomes a limiting factor due to the exponential growth of the number of states.  While we can capture the entire set of attractor states for networks as large as $\sim$300 nodes by leveraging the modular structure of their topology (see {\it Modularity} subsection above), this approach does not directly produce basin sizes.
For smaller networks, we can directly compute them, but for the larger biological networks, this is infeasible.  While we can sample a fixed number of initial conditions to estimate basin sizes, our inability to distinguish attractors that we never find in sampling from those with very small basin size limits the usefulness of this approach. Instead, we will leverage the knowledge of the attractors' identities and the fact that in Eq.~\ref{eq:scaling-corrected} we only need to know the number of them that are isolated. 

In what follows, we present an efficient strategy for identifying the backwards reachable set of states in a Boolean network. We employ this method to quantify the exact number of isolated attractors $s$ within the complete set of attractors $r$ identified using the modular approach. 

Let us refer to a generic Boolean network with $n$ nodes and updating rules as in (\ref{system1_alternate}). We will call $S_{i,b}$ the subset of the configuration space $S$ containing all the states satisfying the non-linear equation 
$f_i (x_1,\dots,x_n)=b$, where $b=0,1$. Therefore, given a state $X = (x_1,\dots,x_n)$, its 
backward reachable set, defined as the set of its pre-images, is
\begin{equation}
S_{X} = S_{1,x_1} \cap S_{2,x_2} \cap \dots \cap S_{n,x_n} \ .
\label{eq:intersection}
\end{equation}
Finding the $2n$ sets $S_{i,b}$ is equivalent to finding the pre-images of each state in $S$. A straightforward algorithm for finding $S_{i,b}$ would evaluate $f_i (x_1,\dots,x_n)$ for all $(x_1,\dots,x_n)\in S $, and select all the states for which $f_i (x_1,\dots,x_n) = b$. This approach would require $2n\times2^n$ operations, as opposed to the $n\times2^n$ operations we would need to determine the entire forward evolution of the network.
In general, using a reverse algorithm is not a better way of solving the dynamics of a network \cite{wuensche1992global}. 
Yet for our problem we only need to find the backward reachable sets for each known fixed point, to determine if its only pre-image is itself and it is therefore an isolated fixed point. From a computational standpoint, additional simplifications arise from the fact that the in-degree of the nodes in the network, i.e., the average number of inputs in the $f_i$ functions,
is often much smaller than $n$. Considering this, a more efficient method for performing the intersection in Eq.~\ref{eq:intersection} is to employ a parametric approach based on conditions solely related to the variables involved in each Boolean function. 

If we call $P_{i,b}$ the set of sets of parametric equations defining the elements of $S_{i,b}$, then the set of sets of equations defining the elements of $S_{X}$ is 
\begin{equation}
P_{X} = P_{1,x_1} \bar\times P_{2,x_2} \bar\times \dots \bar\times P_{n,x_n} \ ,    
\label{products}
\end{equation}
where $\bar\times$ denotes an {\it unordered} and {\it repetition-free} Cartesian product of sets, from which sets containing {\it inconsistent equations have been removed}.

\vspace{4mm}\noindent{\bf Examples:} Providing examples will help clarify our method. One benefit of referencing the parametric sets $P_{i,b}$ is that they serve as a highly concise representation of the sets $S_{i,b}$ when they encompass substantial subsets of the state space. For instance, an updating function with only $k$ inputs will impose constraints solely on the $k$ variables involved, leaving the remaining $n-k$ unconstrained. The number of combinations of these $k$ variables will be multiplied by the $2^{n-k}$ number of combinations of the unconstrained variables. It is not uncommon in the biological networks we study to encounter functions with only one input out of tens of possible genes. In a scenario like this, having just twenty nodes can result in the order of the set $S_{i,b}$ being half a million, but its parametric set $P_{i,b}$ includes only one condition for its input.

Next, let us examine an example of how the products of the $P_{i,b}$ sets are calculated. For simplicity, we will refer to the smallest biological network in our database: a gene regulatory network model for mammalian cortical area development from Ref.~\cite{giacomantonio2010boolean}. This model comprises just five genes: {\it Coup\_fti}, {\it Emx2}, {\it Fgf8}, {\it Sp8}, and {\it Pax6}. The network admits two fixed-point attractors,
$X_1 = [0, 0, 1, 1, 1]$ and $X_2 = [1, 1, 0, 0, 0]$ (neither of them being isolated). Let us find the pre-images of $X_1$. From the updating rules
\begin{eqnarray*}
    Coup\_fti & = &  \overline{ Sp8 {\  \bf OR \ } Fgf8  } \\
    Emx2 & = & Coup\_fti {\ \bf AND \ } \overline{Fgf8} {\ \bf AND \ } \overline{Pax6} {\ \bf AND \ } \overline{Sp8} \\
    Fgf8 &=& Fgf8 {\ \bf AND \ } Sp8 {\ \bf AND \ } \overline{Emx2}  \\
    Sp8 &=&   Fgf8    {\ \bf AND \ } \overline{Emx2}    \\
    Pax6 & = & Sp8 {\ \bf AND \ } \overline{Coup\_fti} {\ \bf AND \ } \overline{Emx2} \ 
\end{eqnarray*}
(where an overline represents the Boolean operator {\ \bf NOT})
we can deduce the parametric sets relevant to $X_1$:

\begin{eqnarray*}
P_{Coup\_fti,0} & = & \{  \{Sp8 = 1\} ,\\
                &   & \{ Fgf8 = 1, Sp8 = 0 \}\} \\
P_{Emx2,0} & = &    \{\{ Coup\_fti = 0 \},\\
            & &
                    \{ Coup\_fti = 1 ,Fgf8 = 1\},\\
         & &    \{ Coup\_fti = 1 ,Fgf8 = 0,Pax6=1\},\\
         & &    \{ Coup\_fti = 1 ,Fgf8 = 0, Sp8=1, Pax6=0\}
\}       \\
P_{Fgf8,1} & = & \{\{ Fgf8 = 1, Emx2 = 0, Sp8 = 1 \}\} \\
P_{Sp8,1} & = & \{\{  Emx2 = 0, Fgf8 = 1 \}\}\\
P_{Pax6,1}  & = & \{\{ Coup\_fti = 0, Emx2 = 0, Sp8 =1 \}\} \ .
\end{eqnarray*}

For example, 
\[
 P_{Coup\_fti,0} \bar{\times} P_{Fgf8,1} = \{\{ Fgf8 = 1, Emx2 = 0, Sp8 = 1 \}\}  
 = P_{Fgf8,1} \ ,
\]
because the products including the contradicting conditions $Sp8 = 0$ and $Sp8 = 1$ have been removed.

Once all five sets are multiplied, the result is
\[
P_{X_1} = \{  \{  Coup\_fti = 0, Emx2 = 0, Fgf8 = 1 , Sp8 = 1  \}   \} \ . 
\]
As no condition is set on $Pax6$,
\[
S_{X_1} = \{ [0,0,1,1,0] ,[0,0,1,1,1]  \} \ .
\]
One of the two pre-images is just $X_1$, as expected from that fact that it is a fixed-point.

For isolated fixed-points, this calculation returns a set of order one, where the only element is the fixed-point itself. Therefore, for each network, we perform this test on all of its fixed-points in order to count the number $s$ of them that are isolated.

In computational terms, each set $P_{i,b}$ is essentially a dataframe. This dataframe has columns representing the input nodes of the function $f_i$, and each row corresponds to a combination of dynamic variables (node dynamical states) for which $f_i$ evaluates to $b_i$. The solution to the problem involves obtaining the result sequentially by taking Cartesian products of all $n$ dataframes and subsequently eliminating inconsistent rows.

A potential complication arises from a temporary combinatorial growth in the number of rows when products are computed between conditions over functions that do not share many inputs. This occurs when there are few cancellations. To address this, we initially conduct a Louvain community detection analysis on the network \cite{blondel2008fast}. We then perform the first round of products among dataframes associated with functions that exhibit the highest overlap among their inputs. Finally, we conduct the ultimate round of products among the results obtained in this manner. (This is computationally more efficient than using the modular structure for the first round of products, as it removes the limitation imposed by having a strict hierarchical structure.) This approach provides an efficient method for determining the backward reachable sets of our candidate isolated points.

\section{Discussion}

A comprehensive understanding of biological control is fundamental for both practical applications, such as genetic reprogramming \cite{MulSch11,kamimoto2023gene}, regenerative medicine \cite{tewary2018stem}, and drug design \cite{ghosh2012network,
fortney2012networx,
lamb2006connectivity}, and for advancing our theoretical understanding of cellular coordination in systems biology \cite{davidson2010regulatory}.
As we explore the intricacies of controlling biological systems, essential questions arise: How easily can we exert control over these systems? What factors in a system's design and behavior contribute to the ease or difficulty of control? 

Alternative approaches to modeling the control of biological networks exist, and results can vary significantly depending on the goals set for the control. For example, the ability to attain any possible target state in a typical GRN described by a linear, continuous system requires control over more than 80\% of the network \cite{liu2011controllability}. 
Here we focus instead on the discrete, nonlinear case, and we use a definition of control that aims to force the system to a single one of its original attractor states from any initial condition. 

With a few notable outliers, our previous work on this type of control indicated that the average required number of control nodes primarily depends on the number of preexisting attractors and shows no dependence on the system's size. While these outliers were not numerous enough to undermine the general trend of `easy controllability',
they raised the question of whether biological systems exist that are inherently harder to control.

This work represents a more in-depth analysis of these outliers. Similarly to Ref.~\cite{borriello2021basis}, we compare the results from our database of biological networks to those obtained in several ensembles of systems with randomly assigned network topologies and dynamic rules.  These random ensembles  are particularly important to include in that they produce a larger number of harder to control networks.  Additionally, we roughly doubled the number of biological networks we tested (increasing from 49 to 104 models) through the inclusion of  networks cataloged in Ref.~\cite{kadelka2020meta}. Importantly, no new outliers from our original scaling law were observed. 

Our primary observation here unveils a consistent pattern in these outlier networks, found in both the biological and random ensembles. This pattern originates from the presence of isolated fixed points—fixed points that attract no other states and prove to be among the attractors most challenging to control, with a control kernel that often includes all or almost all core nodes in the network. 

Why is there a connection between unstable fixed points and difficult control?  One simple way to get both instability and difficult control is exemplified in our random network ensemble that uses threshold updating rules.  When thresholds are set uniformly low, this leads to a bias toward activation: the dynamics tend to push toward states with more 1s than 0s.  States with few or no 1s can sometimes map to themselves, forming isolated fixed points.  These states are also difficult to control because any node that is allowed to be active leads away toward further spreading of activation.  

This type of bias is also seen in the few biological networks that are unusually difficiult to control.  Among the available biological networks, we found two outliers at the 3-sigma level (triangles in Fig.~\ref{fig:ck-size-predictions}A), both related to the ErbB network in breast cell lines: ``SKBR3 Breast Cell Line Long-term ErbB Network'' and ``HCC1954 Breast Cell Line Long-term ErbB Network.'' These networks, which are part of the same study focused on resistance mechanisms in breast cancer treatments \cite{vonBenHen14}, all include isolated fixed points. Upon examining the attractors that are hardest to control (those with the maximum control kernel size), we find they consist of states where most nodes are inactive, except for a few nodes that self-sustain activity. This appears to be due to a highly biased network core where the all-zero state is an isolated fixed point, coupled with peripheral nodes that can independently sustain themselves. This combination increases the number of hard-to-control fixed points and thereby the average control kernel size. It is worth noting that this ability of peripheral nodes to sustain their own activity may be an effect of the algorithm used in the original study to infer the network logic.

Unfortunately, this connection between unstable fixed points and difficult control itself has exceptions. In particular, it is possible to construct networks that are difficult to control without having globally unstable (isolated) fixed points.\footnote{We describe one such counterexample in the SI (also illustrated in the last row of Fig.~\ref{fig:example-networks-table}),  a network dynamic in which a locally unstable fixed point has maximally distant network states that map back to it.  We find only one similar network in the random ensembles that we study, and that network also contains a globally unstable fixed point. Unsurprisingly, we do not find any network resembling such a contrived counterexample among the biological systems.}
Still, by efficiently counting the unstable fixed points of a network, we are able to better explain the variance in control kernel sizes by incorporating the prevalence of these unstable fixed points. This constitutes a useful tool to predict the expected amount of required control over a network based solely on partial information of the attractor landscape.  While determining the number of attractors in a Boolean network is itself a computationally hard problem (\#P-complete \cite{Kos08}), once this is known, our approach avoids the NP-hard problem of evaluating individual control kernels. 

The observation that difficult control is related to unstable fixed points is reassuring for several reasons. Firstly, these fixed points are not a general feature of models with non-deterministic updating rules. Most sources of stochasticity \cite{shmulevich2009deterministic} would push the dynamics toward more stable attractors characterized by larger basins. 
A weak form of stochasticity arises in asynchronous updating.  In this case the identity of fixed-point attractors remains unchanged, but basin sizes can change.  As a result, it is possible that isolated fixed points would no longer be isolated. A stronger form of stochasticity would include bit-flips.  In this case, we expect that isolated fixed points would effectively disappear.
More importantly, the incompatibility of unstable fixed points with stochasticity makes it highly improbable for such states to carry biological significance, i.e., to represent actual cell types. Their persistence would require fine-tuned preservation of an unstable dynamical state.

The biological irrelevance of such unstable fixed points is also significant for another reason. As they reintroduce an explicit dependence on the network size, these cases of difficult control would contradict our main statement that control does not scale with the size of the system \cite{borriello2021basis}, in direct disagreement with recent results in the genetic reprogramming of mammalian cells \cite{MulSch11}.
In this light, we interpret our correction (Equation \ref{eq:scaling-corrected}) as an empirical method for obtaining a more precise estimate of the required control when employing a simplistic, deterministic model of the regulatory dynamics. 
Verifying that our size-independent scaling law remains valid in models of genetic regulation that include stochasticity is needed but beyond the scope of this work. 

More generally, given our results one might hope  that the amount of control necessary to select macroscopic phenotypes could be approximated without knowing details of the microscopic dynamics.

\section{Code and data availability}

The python code used to recreate the analysis and figures,
as well as CSV files containing the relevant data,
are available as a Zenodo repository under the accession code 
10.5281/zenodo.13819680 [https://doi.org/10.5281/zenodo.13819680].   
This code depends on a number of open-source software 
packages: neet, numpy, pandas, and networkx.

\section{Acknowledgments}

The authors acknowledge Research Computing at Arizona State University for providing HPC resources that have contributed to the research results reported in this paper.

\bibliographystyle{unsrt}
\bibliography{bibliography}

\renewcommand{\thetable}{S\arabic{table}}
\renewcommand{\thefigure}{S\arabic{figure}}
\renewcommand{\thesection}{S\arabic{section}}
\setcounter{table}{0}
\setcounter{figure}{0}
\setcounter{section}{0}

\clearpage

\section*{Supplementary information}

\section{Constructed exceptional network}

It is possible to construct a network that has a fixed point with maximal control kernel $|\mathrm{CK}| = n_\mathrm{c}$ without having any isolated fixed points. Here we describe one such construction.

For the first node,
\begin{equation}
    x_1' = \begin{cases} 
    0 & \text{if $\sum_i x_i = 0$ or $n$} \\
    1 & \text{otherwise} \end{cases}
\end{equation}
and for all other nodes $j$, 
\begin{equation}
    x_j' = \begin{cases} 
    0 & \text{if $\sum_i x_i = 0$ or $1$ or $n$} \\
    1 & \text{otherwise} \end{cases}
\end{equation}
These dynamics produce two attractors: one in which all nodes are 0 and one in which all nodes are 0 except for the first.  The control kernel for the second attractor is designed to have size $n$, but neither attractor has a basin size of 1 (specifically, the basin sizes are $n$ and $2^n -n$).

\section{Redundant regulators}
\label{section:redundant}

In this section, we address potential issues arising from discrepancies between the dynamics of a regulatory network and its graph representation. These issues stem from redundant expressions in the updating rules that can make the network present inoperative edges with no causal effect.

Previous work has shown that inoperative edges are rare in the 
biological networks, but not so in random networks \cite{DanKimMoo18}.
We have examined the most apparent inoperative connections in the Cell Collective database and in the ensembles of random networks we generated. Additional networks from Ref.~\cite{kadelka2020meta} have been curated similarly. However, we cannot guarantee that all inoperative edges have been eliminated. We aim to explain why, for the purposes of our study, a more meticulously curated database is not necessary.

Our primary objective is to elucidate why some regulatory networks are unexpectedly difficult to control. In this context, ``difficult'' refers to a significant deviation from the $\langle|\mathrm{CK}|\rangle \sim \log_2(r)$ scaling law in Eq.~\ref{eq:scaling}.

These issues can appear in several forms: inoperative edges, inoperative input nodes, and disconnected graphs. The latter two issues are more visible manifestations of inoperative edges. In the following, we will demonstrate that, in all three cases, these undesirable features do not make the network more difficult to control than anticipated.  It would be possible but nontrivial to remove all inoperative edges before the analysis, so we choose not to do so here because it will not affect our conclusions.


\vspace{4mm}
\noindent {\bf Inoperative edges:} For each network, we evaluate $\langle|\mathrm{CK}|\rangle$ and $r$ by explicitly solving the dynamics. Our analysis therefore does not depend directly on network topology except for our removal of peripheral trees to find the network core and compute $n_c$.

Inoperative edges could result in an overestimate of $n_c$, due to inoperative edges that hide output peripheral trees by making them look like they influence the core.  For a given network, if we compute a number of isolated fixed points $s > 0$ while including possible inoperative edges, then we know that our measurement of $n_c$ is correct: note that extra peripheral nodes created by inoperative edges would lead to basins of attraction with size $> 1$ and therefore would mean that we would measure $s = 0$.  If we fixed this issue by testing for and removing inoperative edges, this would only affect estimated control kernel sizes for cases in which we computed $s=0$ (since the extra edges would have the effect of making it appear that $s=0$ in cases where $s$ is actually greater than 0.) We find empirically that difficult control occurs only in networks for which we computed $s>0$, so removing inoperative edges would not affect our conclusions about difficult control.

\vspace{4mm}
\noindent {\bf Inoperative input nodes:} In cases where a node has neither real input nor output edges, it acts as an inoperative input node. This results in an artificial duplication of the number of attractors, as the inoperative input node introduces two possible entries (0/1) that are not present in the states of a more accurately curated network. Our algorithm for determining $\langle|\mathrm{CK}|\rangle$ will automatically include this irrelevant input, thus increasing $\langle|\mathrm{CK}|\rangle$ by 1, consistent with the $\langle|\mathrm{CK}|\rangle \sim \log_2(r)$ scaling law.

\vspace{4mm}
\noindent {\bf Disconnected graphs:} Inoperative edges might obscure the fact that, from a dynamical perspective, the network can be reduced to two independent dynamical systems. After removing all inoperative edges, the network representation would correspond to two or more disconnected graphs. In this case, the control kernel for each global attractor must include all controlling nodes from the control kernels of the individual, fully connected components. The total number of attractors for the entire system would be the product of the attractors from each component. The $\langle|\mathrm{CK}|\rangle \sim \log_2(r)$ scaling law remains valid due to the product rule of logarithms. (The inoperative input nodes in the previous scenario are the simplest example of disconnected subgraphs consisting of a single node.)

\section{Exploring the shapes of residual distributions}
\label{section:residuals}

In this section, we examine the statistical distributions of our predictions. As in Figure \ref{fig:variance-histograms}B and C, we consider the distribution of the residual differences between the actual $\langle |\textrm{CK}| \rangle$ and our predictions using both our simple logarithmic scaling (Eq.~\ref{eq:scaling}) and our {\it corrected} scaling (Eq.~\ref{eq:scaling-corrected}).

Figures \ref{fig:variance-histograms}B and C provide a visual confirmation that our corrected expressions for $\langle |\textrm{CK}| \rangle$ represent an improvement over the simpler estimates. Specifically, the distribution of the corrected residuals shows greater consistency with zero, with a smaller variance, due to a reduction in the predictions asymmetrically contributing to the right tail of our distribution. Here, we provide an estimate of the improvement this feature offers.

Before adopting any test for deviation from normality, it is important to emphasize that we have no reason to expect the distribution of our predicted residuals to follow a normal distribution, whether using our simpler or corrected model. We can still use such tests to quantify the reduction in the weight of the residual tails relative to any Gaussian fit. 
If our corrected approximation reduces the number of data points in the right tail of the distribution, this would manifest as a significant change in the test statistic of a normality test, e.g. a Shapiro-Wilk test as in the following. We indeed observe this effect. We analyzed the four distributions of residuals that correspond to the choice of biological vs. random networks together with the choice of our simpler vs. corrected scaling. The results of our test are as follows: the Shapiro-Wilk test statistic ($W$) increased from 0.64 to 0.85 for the biological network, and from 0.68 to 0.87 for the random networks.
In both cases the test statistic increases by about $30\%$. (We also observe a corresponding increase in the p-values, which we should disregard, as our goal is not to validate the null hypothesis of normality.)

\begin{figure}
    \centering
    \includegraphics[width=\linewidth]{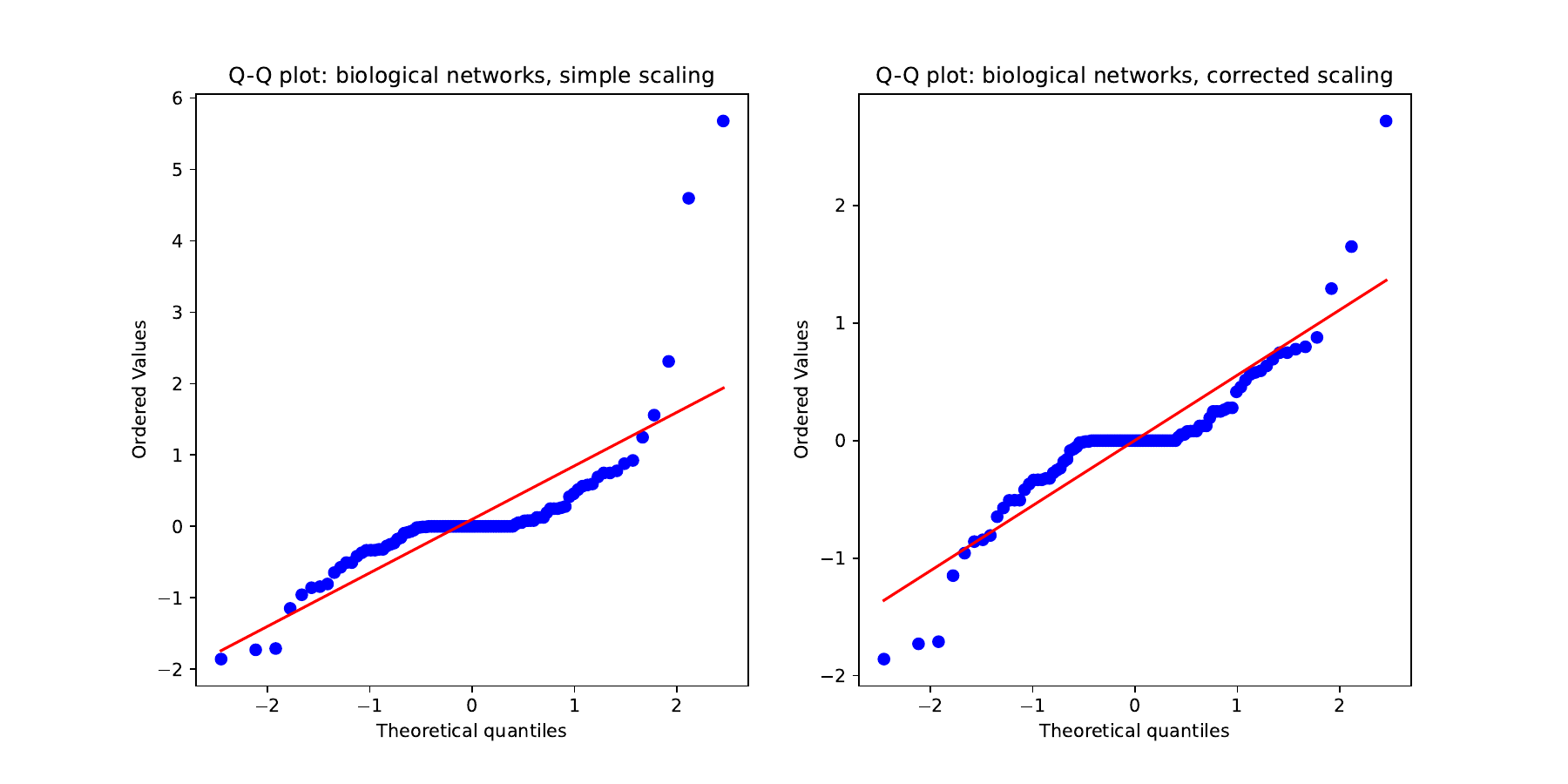}\\
    \includegraphics[width=\linewidth]{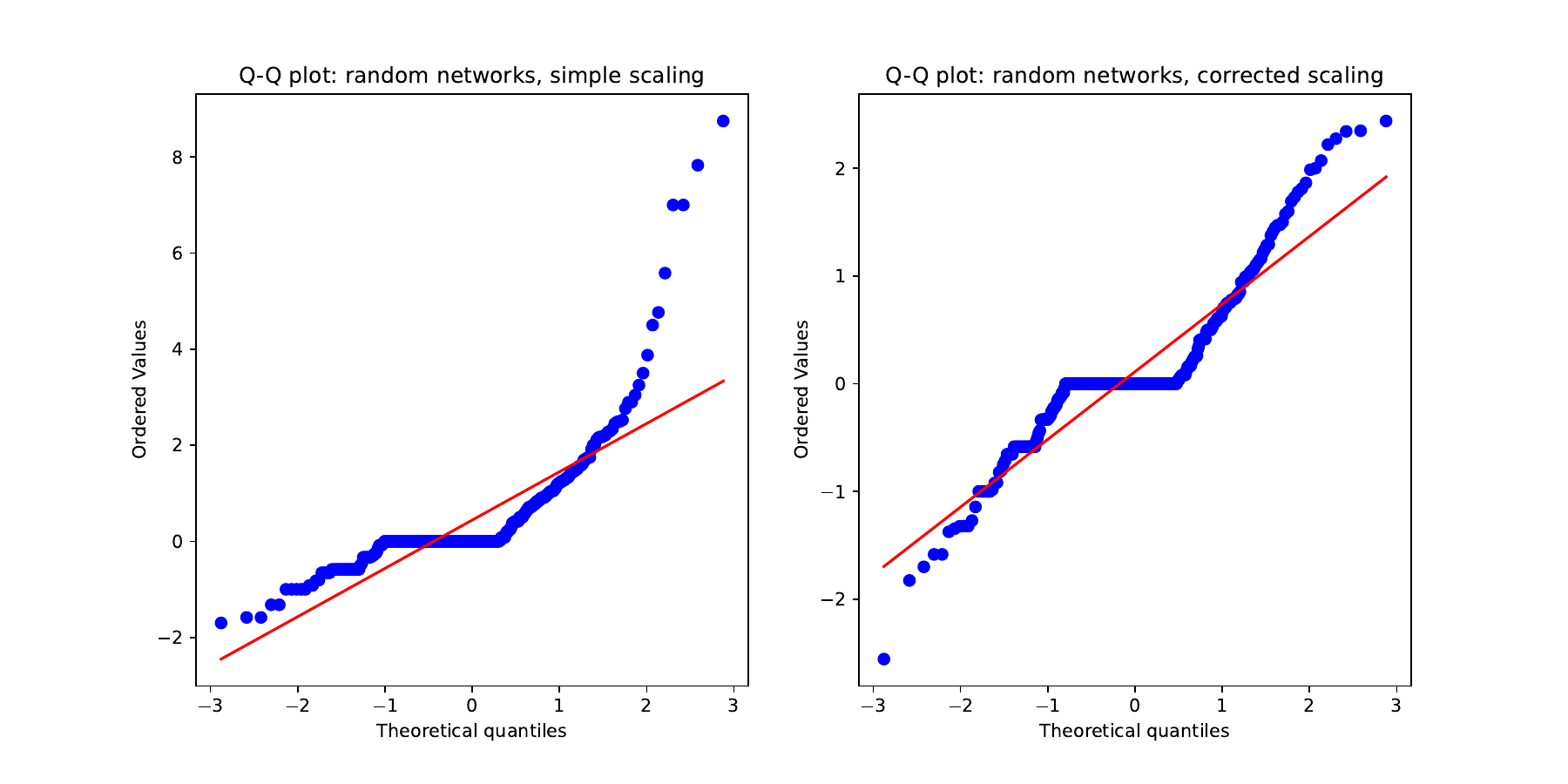}
    \caption{
      Quantile–quantile (Q–Q) plots of the residual distributions shown in Fig. \ref{fig:variance-histograms}B and \ref{fig:variance-histograms}C. In all four cases displayed, the vertical axes ({\it ordered quantiles}) represent the observed residuals, and the horizontal axes ({\it theoretical quantiles}) represent the expected values assuming normally distributed residuals. A perfect match to normality would align with the red diagonal line, indicating linear scaling. The two plots on the left correspond to predictions using logarithmic scaling, while the plots on the right use our corrected scaling. The top row shows results for biological networks, and the bottom row shows results for random networks.
      }
    \label{fig:q-q-plots}
\end{figure}

A more meaningful interpretation comes from complementing these changes in the test statistic with a visual inspection of the quantiles of our distributions vs. the quantiles of a normal distribution. These Q--Q plots, for all four cases discussed here, can be found in Figure~\ref{fig:q-q-plots}, where the vertical axes ({\it ordered quantities}) represent our predictions, and the horizontal axes ({\it theoretical quantities}) show the expectation that would correspond to normally distributed residuals. Perfect normality of our distributions would correspond to a linear scaling of our data with respect to the theoretical quantiles, aligning with the red line shown in each panel.

For both biological and random networks, the dominant source of deviation from normality is represented by the largest ordered values in our distribution. This is the same result already shown in Figure~\ref{fig:variance-histograms}, magnified here by revisiting it in terms of how much departure from a normal distribution they would contribute to if a normal distribution were the expectation.

One notable observation from the Q--Q plots is that the remaining primary source of non-normality in our distributions is the higher frequency with which our predictions are more accurate than a normal distribution would suggest once fitted to accommodate the tails of our distribution of residuals. This is evident in our figures as the horizontal plateau at ordered values equal to zero, where the number of cases with $\langle |\textrm{CK}| \rangle$ approximately zero occurs much more often than expected by a normal distribution. Our corrected approximation preserves this positive feature and decreases the discrepancies at larger observed values by reducing the instances where the simpler scaling underestimates the actual $\langle |\textrm{CK}| \rangle$.

\section{Additional supplemental figures}

\begin{figure}
    \centering
    \includegraphics[width=0.75\linewidth]{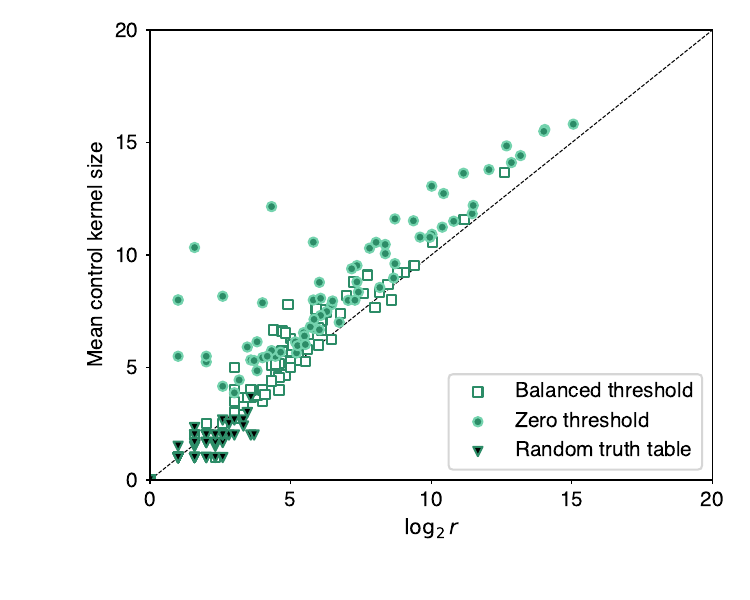}
    \caption{Mean control kernel sizes as a function of the logarithm of the number of attractors for three random network ensembles.  Here we replot Fig.~\ref{fig:ck-size-predictions}B to emphasize differences that arise due to constructing random networks in different ways: using random truth tables (dark triangles), threshold networks with balanced threshold (light squares), and threshold networks with zero threshold (green circles).  Note that all the outliers identified as triangles in Fig.~\ref{fig:ck-size-predictions}B come from the zero threshold ensemble.   This figure omits 4 networks that were included in Fig.~9A of Ref.~\cite{borriello2021basis} due to our more strict threshold that excludes networks that have uncontrollable cycles with large basins.}
    \label{fig:ck-size-random-separate}
\end{figure}

\newgeometry{left=2cm,right=2cm,bottom=2cm,top=2cm}

\begin{figure}
    \centering
    \includegraphics[width=\linewidth]{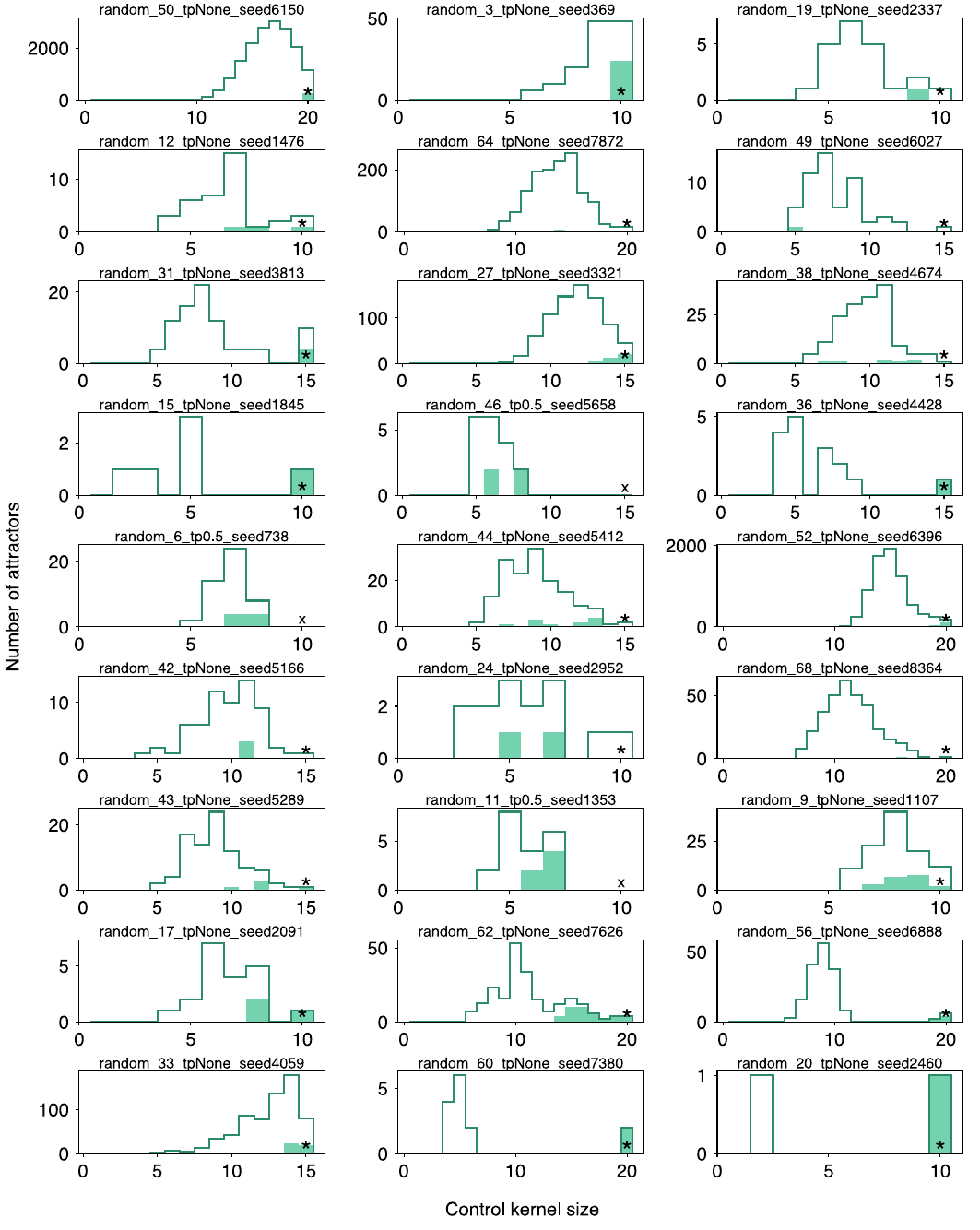}\\
\end{figure}

\begin{figure}
    \centering
    \includegraphics[width=\linewidth]{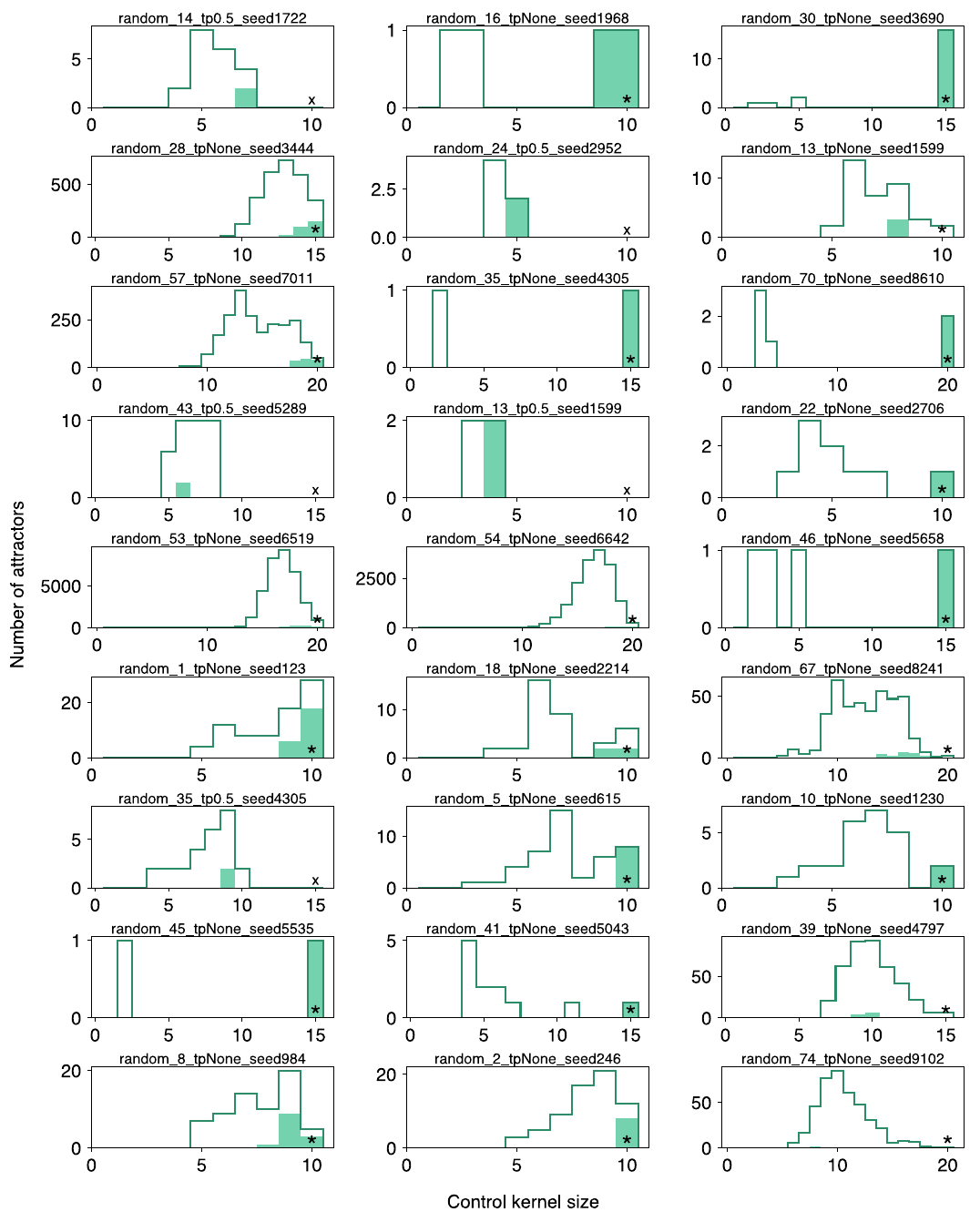}\\
\end{figure}

\begin{figure}
    \centering
    \includegraphics[width=\linewidth]{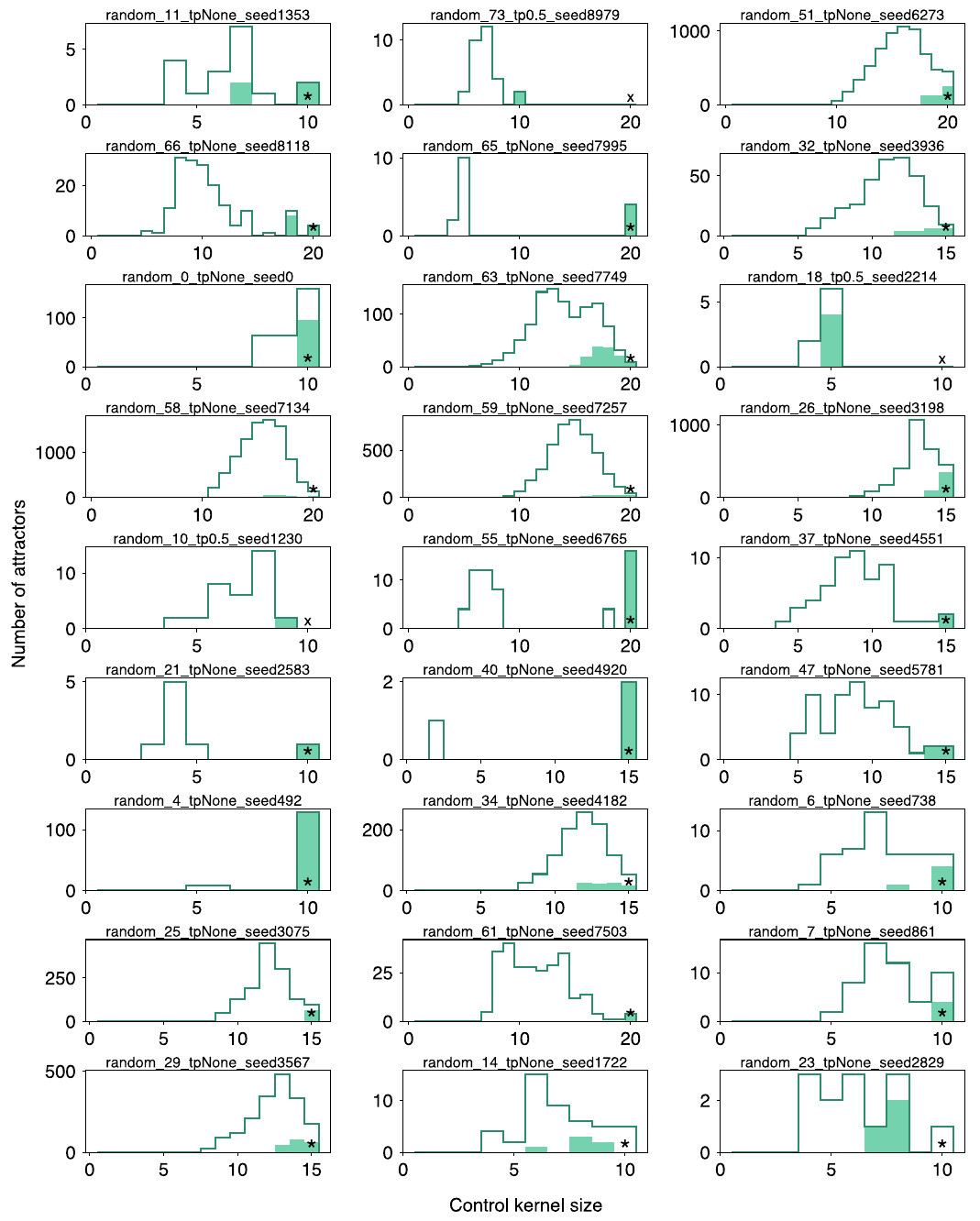}\\
\end{figure}

\begin{figure}
    \centering
    \includegraphics[width=\linewidth]{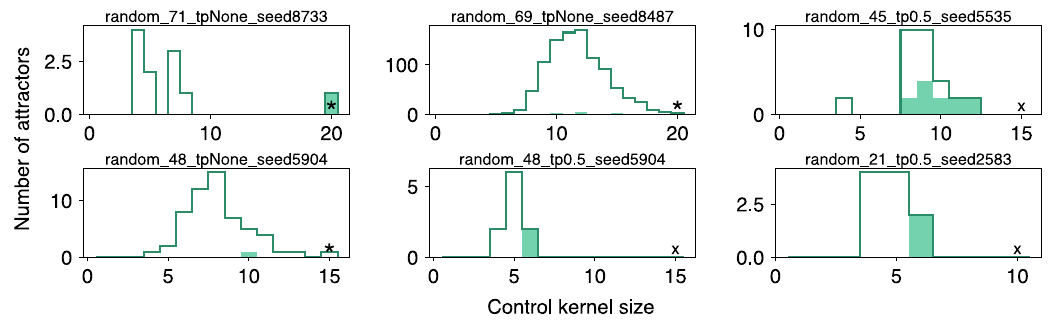}\\
    \caption{\textbf{Histograms of control kernel size for individual random networks with at least one isolated fixed point.} Control kernel sizes of all attractors are displayed with unfilled bars, and the subset corresponding to isolated fixed points are shown as shaded bars.  A star is shown at $n_\mathrm{c}$ for those networks that have at least one control kernel of size $n_\mathrm{c}$ (73 networks), and an X is shown at $n_\mathrm{c}$ otherwise (14 networks).}
    \label{fig:individual-network-histograms-random}
\end{figure}

\begin{figure}
    \centering
    \includegraphics[width=\linewidth]{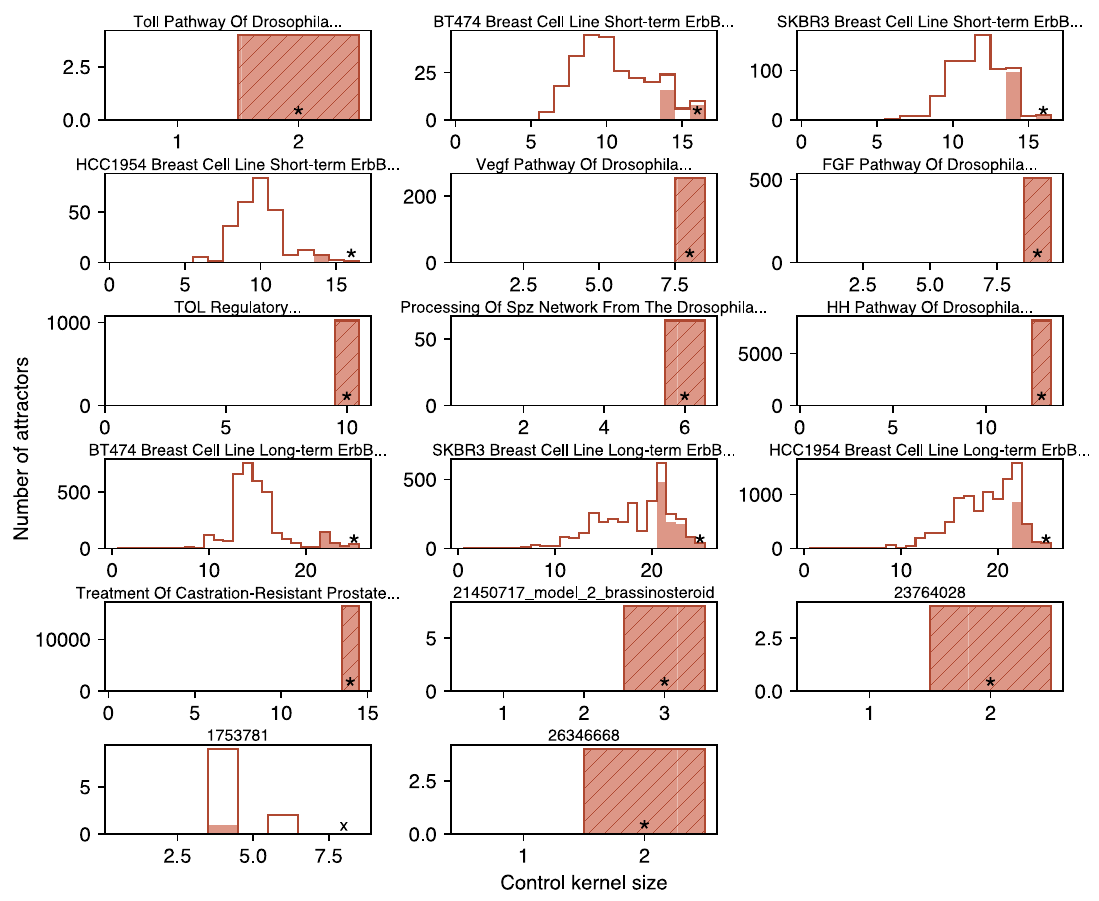}\\
    \caption{\textbf{Histograms of control kernel size for individual biological networks with at least one isolated fixed point.} Control kernel sizes of all attractors are displayed with unfilled bars, and the subset corresponding to isolated fixed points are shown as shaded bars.  Networks without loops have histograms depicted with hatching, as in Fig.~\ref{fig:isolated-and-nonisolated-ck-size-histograms}.  A star is shown at $n_\mathrm{c}$ for those networks that have at least one control kernel of size $n_\mathrm{c}$ (16 networks), and an X is shown at $n_\mathrm{c}$ otherwise (1 network).}
    \label{fig:individual-network-histograms-biological}
\end{figure}

\begin{table}[]
\begin{small}
    \centering
    \begin{tabular}{|l|c|c|c|c|c|c|}

\hline

{\bf  name}	&	{\bf size}	&	{\bf edges}	&	{\bf core nodes}	&	{\bf mean CK size}	&	{\bf attr.}	&	{\bf isol. f.p.}	\\																		
{\bf      }	&	$n$	&	$|E|$	&	$n_\mathrm{c}$	&	$\left<|CK|\right>$	&	$r$	&	$s$	\\						

\hline

Arabidopsis Thaliana Cell Cycl...	&	14	&	66	&	14	&	0	&	1	&	0	\\																		
B Cell Differentiation	&	22	&	44	&	20	&	5.9	&	61	&	0	\\																		
BT474 Breast Cell Line Long-te...	&	25	&	74	&	25	&	14	&	3344	&	256	\\																		
BT474 Breast Cell Line Short-t...	&	16	&	51	&	16	&	9.5	&	253	&	24	\\																		
Body Segmentation In Drosophil...	&	17	&	31	&	12	&	3.4	&	10	&	0	\\																		
Budding Yeast Cell Cycle	&	20	&	46	&	20	&	4.8	&	33	&	0	\\																		
Budding Yeast Cell Cycle 2009	&	18	&	59	&	18	&	0	&	1	&	0	\\																		
CD4+ T Cell Differentiation An...	&	18	&	84	&	18	&	8.6	&	324	&	0	\\																		
Cardiac Development	&	15	&	39	&	15	&	2.8	&	6	&	0	\\																		
Cell Cycle Transcription By Co...	&	9	&	19	&	9	&	1	&	2	&	0	\\																		
Cortical Area Development	&	5	&	14	&	5	&	1	&	2	&	0	\\																		
Death Receptor Signaling	&	28	&	48	&	25	&	4.7	&	97	&	0	\\																		
FGF Pathway Of Drosophila Sign...	&	23	&	33	&	9	&	9	&	512	&	512	\\																		
Fanconi Anemia And Checkpoint ...	&	15	&	66	&	15	&	0	&	1	&	0	\\																		
Guard Cell Abscisic Acid Signa...	&	44	&	82	&	37	&	5.4	&	95	&	0	\\																		
HCC1954 Breast Cell Line Long-...	&	25	&	74	&	25	&	17.8	&	9452	&	1504	\\																		
HCC1954 Breast Cell Line Short...	&	16	&	51	&	16	&	9	&	274	&	10	\\																		
HH Pathway Of Drosophila Signa...	&	24	&	45	&	13	&	13	&	8192	&	8192	\\																		
Human Gonadal Sex Determinatio...	&	19	&	79	&	18	&	2.3	&	3	&	0	\\																		
Iron Acquisition And Oxidative...	&	22	&	40	&	17	&	2	&	4	&	0	\\																		
Lac Operon	&	13	&	25	&	13	&	3.2	&	9	&	0	\\																		
Mammalian Cell Cycle	&	20	&	52	&	18	&	1.7	&	3	&	0	\\																		
Mammalian Cell Cycle 2006	&	10	&	35	&	10	&	0	&	1	&	0	\\																		
Metabolic Interactions In The ...	&	12	&	31	&	11	&	6.2	&	40	&	0	\\																		
Neurotransmitter Signaling Pat...	&	16	&	22	&	14	&	2	&	4	&	0	\\																		
Oxidative Stress Pathway	&	19	&	34	&	17	&	1	&	2	&	0	\\																		
Predicting Variabilities In Ca...	&	15	&	39	&	15	&	2.8	&	6	&	0	\\																		
Pro-inflammatory Tumor Microen...	&	26	&	81	&	26	&	2.5	&	6	&	0	\\																		
Processing Of Spz Network From...	&	24	&	34	&	6	&	6	&	64	&	64	\\																		
Regulation Of The L-arabinose ...	&	13	&	21	&	10	&	4.1	&	21	&	0	\\																		
SKBR3 Breast Cell Line Long-te...	&	25	&	85	&	25	&	17.3	&	3185	&	960	\\																		
SKBR3 Breast Cell Line Short-t...	&	16	&	46	&	16	&	10.7	&	702	&	104	\\																		
Septation Initiation Network	&	31	&	57	&	29	&	9.4	&	2240	&	0	\\																		
T Cell Differentiation	&	23	&	38	&	23	&	5	&	33	&	0	\\																		
T-Cell Signaling 2006	&	40	&	55	&	10	&	3.1	&	10	&	0	\\																		
T-LGL Survival Network 2011 Re...	&	18	&	43	&	17	&	2.3	&	3	&	0	\\																		
TOL Regulatory Network	&	24	&	58	&	10	&	10	&	1024	&	1024	\\																		
Toll Pathway Of Drosophila Sig...	&	11	&	13	&	2	&	2	&	4	&	4	\\																		
Treatment Of Castration-Resist...	&	42	&	65	&	14	&	14	&	16384	&	16384	\\																		
Trichostrongylus Retortaeformi...	&	26	&	59	&	25	&	3.2	&	12	&	0	\\																		
Tumour Cell Invasion And Migra...	&	32	&	158	&	22	&	3.3	&	15	&	0	\\																		
Vegf Pathway Of Drosophila Sig...	&	18	&	26	&	8	&	8	&	256	&	256	\\																		
Wg Pathway Of Drosophila Signa...	&	26	&	43	&	23	&	14	&	16384	&	0	\\																		
Yeast Apoptosis	&	73	&	126	&	30	&	13.3	&	11520	&	0	\\

\hline

\end{tabular} 
    \caption{Cell Collective networks \cite{Helikar2012}.}
    \label{tab:cc_nets}
\end{small}
\end{table}

\begin{table}[]
\begin{small}
    \centering
         \begin{tabular}{|l|c|c|c|c|c|c|}

\hline

{\bf  PubMed ID} & {\bf size}& {\bf edges} & {\bf core nodes} & {\bf mean CK size}   & {\bf attr.} & {\bf isol. f.p.} \\ 
{\bf      } & $n$   &  $|E|$   & $n_\mathrm{c}$   &  $\left<|CK|\right>$ & $r$         & $s$              \\ \hline

1753781	&	8	&	30	&	8	&	3.363636364	&	11	&	1	\\																								
11082279	&	11	&	20	&	11	&	2	&	4	&	0	\\																								
19524598	&	12	&	26	&	12	&	2	&	4	&	0	\\																								
20169167	&	7	&	12	&	7	&	1	&	2	&	0	\\																								
20659480	&	20	&	30	&	13	&	1.666666667	&	4	&	0	\\																								
21563979	&	13	&	25	&	13	&	3.222222222	&	9	&	0	\\																								
21639591	&	25	&	57	&	25	&	2	&	4	&	0	\\																								
21853041	&	11	&	30	&	11	&	2.666666667	&	8	&	0	\\																								
22192526	&	9	&	19	&	8	&	3	&	7	&	0	\\																								
23056457	&	17	&	41	&	16	&	3.833333333	&	12	&	0	\\																								
23469179	&	10	&	23	&	9	&	3.125	&	8	&	0	\\																								
23764028	&	5	&	6	&	2	&	2	&	4	&	4	\\																								
24376455	&	10	&	24	&	9	&	2	&	4	&	0	\\																								
25163068	&	30	&	44	&	27	&	7.265306122	&	241	&	0	\\																								
25967891	&	13	&	42	&	12	&	4.1	&	10	&	0	\\																								
26244885	&	30	&	54	&	29	&	8.397435897	&	1104	&	0	\\																								
26346668	&	30	&	42	&	2	&	2	&	4	&	4	\\																								
27613445	&	25	&	68	&	23	&	3.6	&	22	&	0	\\																								
28187161	&	30	&	54	&	27	&	6.823529412	&	220	&	0	\\																								
28209158	&	9	&	34	&	9	&	1.666666667	&	3	&	0	\\																								
28455685	&	15	&	24	&	15	&	2	&	4	&	0	\\																								
29186334	&	29	&	83	&	16	&	5.238095238	&	25	&	0	\\																								
29596489	&	23	&	55	&	18	&	2.25	&	5	&	0	\\																								
30024932	&	21	&	62	&	16	&	2.333333333	&	6	&	0	\\																								
30104572	&	5	&	12	&	5	&	2	&	3	&	0	\\																								
30148917	&	18	&	30	&	14	&	4	&	16	&	0	\\																								
30323768	&	21	&	36	&	16	&	3.8	&	18	&	0	\\																								
30530226	&	23	&	39	&	19	&	2	&	5	&	0	\\																								
30546316	&	30	&	58	&	21	&	10.60769231	&	1056	&	0	\\																								
31949240	&	24	&	46	&	21	&	5.43902439	&	45	&	0	\\																								
32054948	&	27	&	80	&	24	&	2.25	&	6	&	0	\\																								
19622164\_TGF\_alpha	&	17	&	68	&	9	&	2.8	&	7	&	0	\\																								
19622164\_TGF\_beta1	&	17	&	68	&	9	&	2.8	&	7	&	0	\\																								
20221256\_manual\_mod	&	11	&	27	&	11	&	3.615384615	&	22	&	0	\\																								
21450717\_model\_2\_brassinosteroid	&	10	&	12	&	3	&	3	&	8	&	8	\\																								
21450717\_model\_4\_2\_BRX\_integration	&	20	&	30	&	17	&	3	&	14	&	0	\\																								
21450717\_model\_5\_1\_BRX\_integration	&	20	&	31	&	17	&	2	&	4	&	0	\\																								
21450717\_model\_5\_2	&	20	&	31	&	17	&	2	&	4	&	0	\\																								
23169817\_high\_dna\_damage	&	16	&	60	&	16	&	0	&	1	&	0	\\																								
23169817\_low\_dna\_damage	&	16	&	60	&	16	&	0	&	1	&	0	\\																								
23630177\_metastatic\_melanoma	&	10	&	25	&	9	&	2	&	4	&	0	\\																								
23658556\_model\_1	&	14	&	35	&	14	&	3.777777778	&	10	&	0	\\																								
23658556\_model\_2	&	16	&	38	&	16	&	3.2	&	9	&	0	\\																								
23658556\_model\_3	&	15	&	38	&	15	&	3.75	&	9	&	0	\\																								
23658556\_model\_4	&	17	&	42	&	17	&	3	&	10	&	0	\\																								
23658556\_model\_5	&	15	&	37	&	15	&	3.125	&	8	&	0	\\																								
23658556\_model\_6	&	15	&	37	&	15	&	3.125	&	8	&	0	\\																								
23658556\_model\_7	&	17	&	40	&	17	&	4.5	&	14	&	0	\\																								
23658556\_model\_8	&	16	&	38	&	16	&	4.1	&	12	&	0	\\																								
23658556\_model\_9	&	17	&	42	&	17	&	3.25	&	9	&	0	\\																								
24564942\_threshold	&	8	&	23	&	8	&	2.6	&	5	&	0	\\																								
28426669\_ARF10\_greater\_ARF5	&	16	&	39	&	15	&	3.3	&	14	&	0	\\																								
28426669\_ARF10\_smaller\_ARF5	&	16	&	39	&	15	&	3.3	&	14	&	0	\\

\hline

\end{tabular}
 
    \caption{Additional biological networks from  \cite{kadelka2020meta}.}
    \label{tab:iowa_nets}
    \end{small}

\end{table}

\end{document}